\pdfoutput=1
\documentclass[11pt,twoside,a4paper,cmspaper,final,collab]{cms-tdr}
\def\svnVersion{06f46b5}\def\svnDate{2025/04/02}\def\cmsCernNoTag{CERN-EP-2024-250}\def\cmsCernDate{\today}\def\cmsMessage{Published in the Journal of High Energy Physics as \href{http://dx.doi.org/10.1007/JHEP02(2025)199}{\doi{10.1007/JHEP02(2025)199}.}}
\begin{document}\cmsNoteHeader{B2G-23-004}

\providecommand{\cmsTable}[1]{\resizebox{\textwidth}{!}{#1}}
\newlength\cmsTabSkip\setlength{\cmsTabSkip}{1ex}
\providecommand{\cmsLeft}{left\xspace}
\providecommand{\cmsRight}{right\xspace}
\newcommand{\PR}{\ensuremath{\HepParticle{R}{}{}}\xspace}
\newcommand{\MR}{\ensuremath{{m}_{\PR}}\xspace}
\newcommand{\mja}{\ensuremath{m_{\text{ja}}}\xspace}
\newcommand{\mjb}{\ensuremath{m_{\text{jb}}}\xspace}
\newcommand{\mjc}{\ensuremath{m_{\text{jc}}}\xspace}
\newcommand{\sja}{\ensuremath{s_{\text{ja}}}\xspace}
\newcommand{\sjb}{\ensuremath{s_{\text{jb}}}\xspace}
\newcommand{\mSR}{\ensuremath{m_{\text{85}}}\xspace}
\newcommand{\mCR}{\ensuremath{m_{\text{90}}}\xspace}
\newcommand{\gKK}{\ensuremath{\Pg_{\text{KK}}}\xspace}
\newcommand{\MgKK}{\ensuremath{{m}_{\gKK}}\xspace}
\newcommand{\mgKK}{\ensuremath{{m}_{\gKK}}\xspace}
\newcommand{\mjj}{\ensuremath{m_{\mathrm{jj}}}\xspace}
\newcommand{\Mjjj}{\ensuremath{m_{\mathrm{jjj}}}\xspace}
\newcommand{\mjjj}{\ensuremath{m_{\mathrm{jjj}}}\xspace}
\newcommand{\mjjs}{\ensuremath{m^*_{\mathrm{jj}}}\xspace}
\newcommand{\mjjjs}{\ensuremath{m^*_{\mathrm{jjj}}}\xspace}
\newcommand{\Wjets}{\ensuremath{\PW\text{+jets}}\xspace}
\newcommand{\Zjets}{\ensuremath{\PZ\text{+jets}}\xspace}
\newcommand{\PNetW}{\ensuremath{\text{PNet}_\PW}\xspace}

\cmsNoteHeader{B2G-23-004}

\title{Search for high-mass resonances in a final state comprising a gluon and two hadronically decaying \texorpdfstring{\PW}{W} bosons in proton-proton collisions at \texorpdfstring{$\sqrt{s} = 13\TeV$}{sqrt(s) = 13 TeV}}

\date{\today}

\abstract{A search for high-mass resonances decaying into a gluon, \Pg, and two \PW bosons is presented. A Kaluza--Klein gluon, \gKK, decaying in cascade via a scalar radion \PR, $\gKK\to\Pg\PR\to\Pg\PW\PW$, is considered. The final state studied consists of three large-radius jets, two of which contain the products of hadronically decaying \PW bosons, and the third one the hadronization products of the gluon. The analysis is performed using proton-proton collision data at $\sqrt{s}=13\TeV$ collected by the CMS experiment at the CERN LHC during 2016--2018, corresponding to an integrated luminosity of 138\fbinv. The masses of the \gKK and \PR candidates are reconstructed as trijet and dijet masses, respectively.These are used for event categorization and signal extraction. No excess of data events above the standard model background expectation is observed. Upper limits are set on the product of the \gKK production cross section and its branching fraction via a radion \PR to $\Pg\PW\PW$. This is the first analysis examining the resonant $\PW\PW$+jet signature and setting limits on the two resonance masses in an extended warped extra-dimensional model.}

\hypersetup{
pdfauthor={CMS Collaboration},
pdftitle={Search for high-mass resonances in a final state comprising a gluon and two hadronically decaying W bosons in proton-proton collisions at sqrt(s) = 13 TeV},
pdfsubject={CMS},
pdfkeywords={CMS,triboson,extended warped extra-dimensional models}}
\maketitle 

\section{Introduction}

The search for extensions of the standard model (SM) remains one of the most important goals of the CERN LHC research program. 
Beyond-the-SM (BSM) physics can manifest itself through the production of a new heavy particle that decays into SM particles.
Numerous searches for massive resonances decaying into a pair of SM particles, such as dijet or diboson searches \cite{CMS:2021itu,CMS:2022pjv,CMS:2021klu,CMS:2019gwf,CMS-PAS-B2G-23-008,CMS:2021xor,CMS:2021fyk,CMS:2024phk,ATLAS:2020qiz,ATLAS:2020fry,ATLAS:2019nat,ATLAS:2024qvg,ATLAS:2022zuc,ATLAS:2022enb,ATLAS:2019fgd}, have been conducted without resulting in any compelling evidence for BSM physics.

Several BSM theories can also give rise to resonances resulting in signatures with three or more SM bosons.
Extended models with warped extra dimensions~\cite{Agashe:2016rle,Agashe:2016kfr,Agashe:2017wss,Agashe:2018leo,Kuang:2015gka,Ren:2014bza} are among those that predict such a signal.
These models assume a Randall--Sundrum scenario with an extended bulk space containing two branes in addition to the one on which the SM resides~\cite{Agashe:2016rle}.
Each additional brane introduces a bulk scalar boson, the radion \PR, which stabilizes the size of the extra dimension through the potential of a modulus field~\cite{Goldberger:1999uk}.
One of the main motivations of these models is to tackle the SM hierarchy problem~\cite{Weinberg:1975gm,Susskind:1978ms}, where quantum corrections to the Higgs boson mass need to be fine-tuned to the ratio $(M_{\text{EW}}/M_{\text{Pl}})^2$ in order to maintain the Higgs boson mass at the observed electroweak scale, with $M_{\text{EW}}$ and $M_{\text{Pl}}$ representing the electroweak and Planck scales, respectively. Extended warped extra-dimensional models address the $M_{\text{Pl}}$--$M_{\text{EW}}$ gap by proposing that it is an artifact of the SM not residing in the full bulk space, while simultaneously addressing flavor hierarchy problems.
Such BSM signatures would, however, remain undetected in the above-mentioned dijet and diboson resonance searches since their final states are not fully reconstructed.
Therefore, dedicated searches are needed.

Previous searches for triboson and trigluon signatures resulting from a two-step cascade decay of a heavy resonance have been performed by the CMS experiment~\cite{CMS:2022lqh,CMS:2021mjl,CMS:2022tqn,CMS:2023tep}.
A scenario containing a \PW boson plus two gluons (\Pg) has been probed by the ATLAS experiment~\cite{ATLAS:2022ddh}.
Despite the constraints imposed by these searches, a large fraction of the theoretical parameter space remains unexplored.
In this paper, we consider the specific extended warped extra-dimensional scenario in which only the quantum chromodynamic (QCD) gauge field can propagate into the extended bulk, giving rise to a Kaluza--Klein (KK) gluon \gKK.
The \gKK can decay in a cascade through the lightest radion of the theory.
In a certain parameter space of the model, the dominant decay mode is $\gKK\to\Pg\PR\to\Pg\PW\PW$, which has not yet been directly probed, and is the focus of this analysis.

\begin{figure}[ht!]\centering
\includegraphics[width=0.5\linewidth]{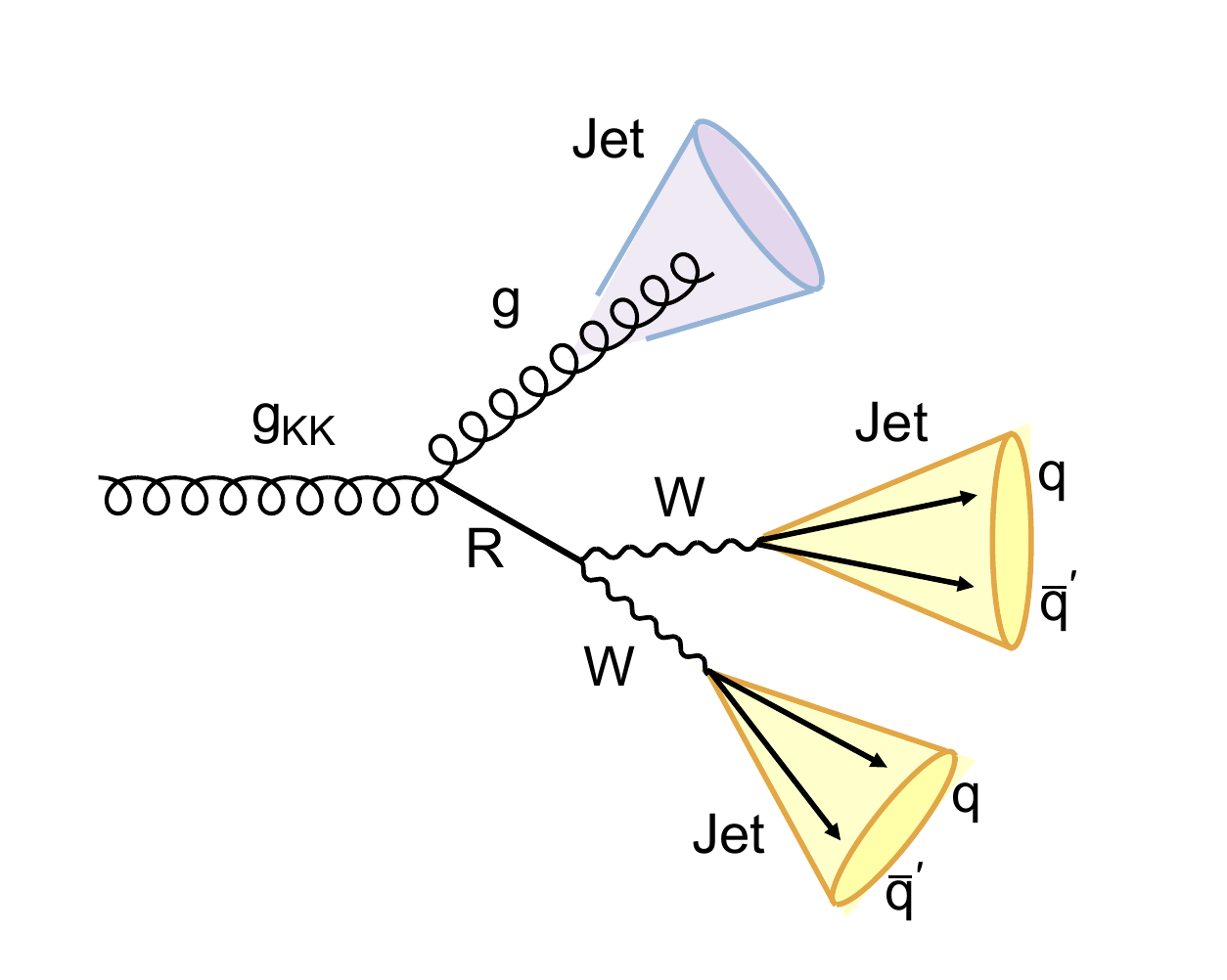}
\caption{A schematic diagram of the decay of a \gKK boson via a radion \PR to the final state considered in this analysis.}
\label{fig:diagram}
\end{figure}

In this paper, we probe the $\Pg\PW\PW$ signature performing a search for BSM physics in the trijet final state. 
Two of the jets, both arising from the posited \PR decay, are identified using jet substructure techniques as originating from hadronic decays of Lorentz-boosted \PW bosons.
No particular requirements are imposed on the third jet, which, in this model, is expected to originate from the gluon.
This choice preserves search generality as it allows other potential signals to be selected.
Figure \ref{fig:diagram} displays a schematic diagram of the signal process, including the cascade decay and the resulting final state probed in this paper.
The overall \gKK mass range explored is 1.25--4.5\TeV.
Within this range, the phase space with a ratio of \PR and \gKK masses of $0.04<\MR/\MgKK<0.9$ is considered.
For $\MR/\MgKK<0.2$ and $\MR/\MgKK>0.8$, the analysis sensitivity is reduced because a significant number of events result in a dijet final state that is not selected.
In the former case, the two \PW bosons are collimated, while in the latter case, the final-state gluon carries relatively little energy.

The analysis is based on proton-proton ($\Pp\Pp$) collision data collected by the CMS experiment during 2016--2018 at $\sqrt{s}=13\TeV$, with integrated luminosities of 36.3~\cite{Sirunyan:2021qkt}, 41.5~\cite{CMS-PAS-LUM-17-004}, and 59.7~\cite{CMS-PAS-LUM-18-002}\fbinv for 2016, 2017, and 2018, respectively, yielding a total of 138\fbinv.
The targeted signal is reconstructed by identifying the two \PW boson jet candidates, whose combined invariant mass \mjj yields the posited \MR. 
These jets, along with a third one, are used to reconstruct the \gKK resonance candidate and compute its invariant trijet mass \Mjjj.
To increase analysis sensitivity, \mjj and \Mjjj are corrected, improving their resolution.
Machine-learning techniques are employed to reconstruct and identify the jets originating from a \PW boson decay.
In previous analyses~\cite{CMS:2022lqh,CMS:2021mjl}, the {DeepAK8} algorithm~\cite{JME-18-002} has been used for this purpose.
In this paper, we use the ParticleNet (PNet) algorithm~\cite{Qu:2019gqs}, which yields a significantly better signal selection efficiency for the same background rejection and exhibits a reduced dependency on the jet mass.
Additionally, we split the selected event sample based on the ratio $\mjj/\Mjjj$.

While this search is optimized for a $\Pg\PW\PW$ signal, it is also sensitive to other signals containing a pair of Lorentz-boosted \PW or \PZ bosons.
In particular, this includes any heavy resonance decaying, with or without an intermediate particle, into $\PX\PW\PW$, $\PX\PW\PZ$, or $\PX\PZ\PZ$, where \PX is a particle resulting in an energetic jet, such as a light or heavy quark, a vector boson, or a BSM particle.

This paper is organized as follows:
Section~\ref{sec:cms} provides a description of the CMS detector.
Section~\ref{sec:Sample} describes the simulated samples used in the analysis.
The triggers employed for data collection and the event reconstruction are discussed in Section~\ref{sec:Objects}.
The event selection and categorization are presented in Section~\ref{sec:event_selection}.
Section~\ref{sec:bkg_estimation} describes the estimation of the SM background.
Systematic uncertainties are discussed in Section~\ref{sec:uncertainties}.
The results and their interpretation are given in Section~\ref{sec:Results}.
The paper is summarized in Section~\ref{sec:sum}.
Tabulated results for this analysis are provided in the HEPData record~\cite{hepdata}.

\section{The CMS detector} \label{sec:cms}

The central feature of the CMS detector is a superconducting solenoid of 6\unit{m} internal diameter, providing a magnetic field of 3.8\unit{T}.
A silicon pixel and strip tracker, a lead tungstate crystal electromagnetic calorimeter (ECAL), and a brass and scintillator hadron calorimeter (HCAL), each composed of a barrel and two endcap sections resides within the solenoid volume.
Forward calorimeters extend the coverage provided by the barrel and endcap detectors up to pseudorapidities of $\abs{\eta} = 5$.
Muons are measured in gas-ionization detectors embedded in the steel flux-return yoke outside the solenoid.

The trijet events of interest are selected using a two-tiered trigger system.
The first level, composed of custom hardware processors, uses information from the calorimeters and muon detectors to select events at a rate of around 100\unit{kHz}, making a decision within the fixed period of about 4\mus following the beam crossing, allowed by the latency implemented in the readout path~\cite{Sirunyan:2020zal}.
The second level, known as the high-level trigger, consists of a farm of processors running a version of the full event reconstruction software optimized for fast processing,
and reduces the event rate to around 1\unit{kHz} before data storage~\cite{Khachatryan:2016bia}.
A more detailed description of the CMS detector, together with a definition of the coordinate system and kinematic variables, can be found in Ref.~\cite{Chatrchyan:2008zzk,CMS:2023gfb}.

\section{Simulated samples} \label{sec:Sample}

The signal is simulated at leading order (LO) using the \MGvATNLO 2.4.2 generator~\cite{MGatNLO},
covering a wide range of \gKK and \PR masses (\MgKK from 1.25 to 4.5\TeV and \MR from 4 up to 90\% of \MgKK).
The corresponding KK gauge coupling values that are used for the interpretation of this result are $g_{\gamma_\mathrm{KK}}=2.65$, $g_{W_\mathrm{KK}}=3$, $g_{{g_\mathrm{KK}}}=6$, and $g_\mathrm{grav}=5$, as proposed in Ref.~\cite{Agashe:2016kfr}.
The widths of both \gKK and \PR are less than 2\% of their masses, thus satisfying the narrow-width approximation.

The dominant SM background originates from jets produced through the strong interaction.
Such events are referred to as QCD multijet events.
While this background is estimated using data, the procedure is tested using simulation.
The QCD multijet events, along with the subdominant background from \Wjets and the \Zjets production, designated as ``V+jets'', are simulated at LO with \MGvATNLO, and matched to parton showers with the MLM~\cite{Alwall:2007fs} algorithm.
The top quark-antiquark pair (\ttbar) and single top quark (\PQt) background production are modeled at next-to-LO (NLO) in QCD using the \POWHEG 2.0 generator~\cite{POWHEG1,POWHEG2,POWHEG3,Alioli:2011as,Alioli:2009je,Frederix:2012dh}.
The minor diboson background is simulated at NLO with either \POWHEG ($\PW\PW$ production) or \MGvATNLO ($\PW\PZ$, $\PZ\PZ$), collectively referred to as ``VV''.
Contributions from other processes are found to be negligible.

The signal and background samples for all three years of data taking are generated with the LO or next-to-NLO NNPDF3.1 parton distribution functions (PDFs)~\cite{nnpdf}, ensuring that the order of the PDFs matches that of the matrix element calculations.
The parton showering, fragmentation, and hadronization for all samples is performed using \PYTHIA 8.230~\cite{Sjostrand:2014zea} with the underlying event tune CP5~\cite{Sirunyan:2019dfx}.
The CMS detector response is modeled using the \GEANTfour package~\cite{Agostinelli:2002hh}.
The simulated events include the contribution of particles from additional $\Pp\Pp$ 
interactions within the same or neighboring bunch crossings, referred to as pileup (PU), and are corrected to reproduce the distribution of the number of PU interactions observed in the data.

\section{Event reconstruction} \label{sec:Objects}

The primary interaction vertex (PV) is taken to be the vertex corresponding to the hardest scattering in the event, evaluated using tracking information alone, as described in Section 9.4.1 of Ref.~\cite{CMS-TDR-15-02}.

The particle-flow (PF) algorithm~\cite{CMS-PRF-14-001} aims to reconstruct and identify each interacting particle in an event, with an optimized combination of information from the various elements of the CMS detector.
The energy of electrons is determined from a combination of the track momentum at the PV, the corresponding ECAL cluster energy, and the energy sum of all bremsstrahlung photons attached to the track.
The energy of muons is obtained from the curvature of the corresponding track.
The energy of charged hadrons is determined from a combination of their momentum measured in the tracker and the matching ECAL and HCAL energy deposits,
corrected for the response function of the calorimeters to hadronic showers.
Finally, the energy of neutral hadrons is obtained from the corresponding corrected ECAL and HCAL energies.

For each event, hadronic jets are clustered from these reconstructed particles using the infrared- and collinear-safe anti-\kt jet finding algorithm~\cite{Cacciari:2008gp,Cacciari:2011ma} as implemented in the \textsc{FastJet} package~\cite{Cacciari:2011ma}.
The clustering algorithm is run twice over the same inputs with the distance parameters 0.4 and 0.8, resulting in AK4 and AK8 jet collections, respectively.
Jet momentum is determined as the vectorial sum of all particle momenta in the jet, and is found from simulation to be, on average, within 5 to 10\% of the true momentum over the entire transverse momentum spectrum and detector acceptance~\cite{Khachatryan:2016kdb}.

The PU per particle identification algorithm~\cite{Sirunyan:2020foa,Bertolini:2014bba} is used to mitigate the effect of PU in AK8 jets at the reconstructed-particle level, making use of local shape information, event PU properties, and tracking information.
A local shape variable is defined, which distinguishes between collinear and soft diffuse distributions of other particles surrounding the particle under consideration. 
The former is attributed to particles originating from the hard scattering and the latter to particles originating from PU interactions. 
Charged particles identified to be originating from PU vertices are discarded. 
For each neutral particle, the local shape variable is computed using the surrounding charged particles compatible with the PV within the tracker acceptance ($\abs{\eta} < 2.5$), and using both charged and neutral particles in the region outside of the tracker coverage. 
The momenta of the neutral particles are then rescaled according to their probability to originate from the PV deduced from the local shape variable, thus superseding the need for jet-based PU corrections~\cite{Sirunyan:2020foa}.

Jet energy corrections are derived from simulation to bring the measured response of jets to that of particle-level jets on average.
In situ measurements of the momentum balance in dijet, photon+jet, \PZ{}+jet, and multijet events are used to account for any residual differences in the jet energy scale (JES) between data and simulation~\cite{Khachatryan:2016kdb}.
The jet energy resolution (JER) measured using a dijet balance technique amounts typically to 15--20\% at 30\GeV, 10\% at 100\GeV, and 5\% at 1\TeV~\cite{Khachatryan:2016kdb}.
Additional selection criteria are applied to each jet to remove jets potentially dominated by anomalous contributions from various subdetector components or reconstruction failures~\cite{CMS-PAS-JME-16-003}.

The AK4 jets originating from the hadronization of bottom (\PQb) quarks are identified (``\PQb tagged'') using a deep neural network algorithm \textsc{DeepJet}~\cite{Bols:2020bkb,CMS-DP-2018-058} that takes as input: tracks within the jet that are displaced from the PV, identified secondary vertices, jet kinematic variables, neutral PF candidates, and information related to the presence of soft leptons in the jet.
A threshold on the {DeepJet} score is used, corresponding to a 0.1\% probability of misidentifying a light-flavor quark or a gluon ($\PQu\PQd\PQs\Pg$) AK4 jet with $\pt > 30\GeV$ and $\abs{\eta}<2.5$ as a \PQb quark jet, while maintaining a 67\% signal efficiency.

The missing transverse momentum vector, \ptvecmiss, is computed as the negative vector sum of the transverse momenta ($\Vec{\pt}$) of all the PF candidates in an event, and its magnitude is denoted as \ptmiss~\cite{Sirunyan:2019kia}.
The \ptvecmiss is modified to account for corrections to the energy scale of the reconstructed jets in the event.
Anomalous high-\ptmiss events can occur due to a variety of reconstruction failures, detector malfunctions, or noncollision backgrounds.
Such events are rejected by conditions that are designed to identify more than 85--90\% of the spurious high-\ptmiss events with a single efficiency loss of less than 0.1\%~\cite{Sirunyan:2019kia}.

The AK8 jets containing hadronic decays of \PW bosons are identified using the PNet machine learning algorithm~\cite{Qu:2019gqs}, with PF candidates and secondary vertices as input.
The algorithm output classifies jets by assigning one score for each class trained.
The algorithm is used in its mass-decorrelated version~\cite{Louppe:2016ylz}.
For the identification of \PW bosons, the ratio of the PNet scores of the $\PW\to\PQq\overline{\PQq}'$ class to the sum of all QCD multijet classes is formed and referred to as \PNetW in the following.
We additionally make use of the groomed jet mass ($m_\text{j}$) for further discrimination of \PW boson jets from background jets.
The groomed jet mass is calculated after applying a modified mass-drop algorithm~\cite{Dasgupta:2013ihk,Butterworth:2008iy} to AK8 jets, with parameters $\beta = 0$, $z_\text{cut} = 0.1$, and $R_0 = 0.8$ also known as the \textit{soft drop} algorithm~\cite{Larkoski:2014wba}. 
The variables are calibrated in a \ttbar enriched sample with hadronically decaying \PW boson~\cite{Khachatryan:2014vla}.
Further details on the calibration method used for this analysis are given in Section~\ref{sec:bkg_estimation}.
The soft-drop mass is meant when referring to the jet mass in the following.

Isolated muon and electron candidates, collectively referred to as leptons \Pell in the following, are reconstructed as follows in order to veto events containing them.
Muon candidates are required to fulfill $\pt>20\GeV$, to be within the geometrical acceptance of the muon detectors ($\abs{\eta}<2.4$), and are reconstructed by combining the information from the silicon tracker and the muon chambers~\cite{Sirunyan:2018fpa}.
Electron candidates with $\pt>35\GeV$ and within $\abs{\eta}<2.5$ are reconstructed using an algorithm that associates fitted tracks in the silicon tracker with electromagnetic energy clusters in the ECAL~\cite{Khachatryan:2015hwa}.
Electron candidates identified as coming from photon conversions in the detector are also rejected.
Identified muons and electrons are required to be isolated from hadronic activity in the event.
The isolation variable is defined by summing the \pt of all the PF candidates in a cone of radius ${\Delta R = \sqrt{\smash[b]{(\Delta\eta)^{2}+(\Delta\phi)^{2}}} = 0.4\,(0.3)}$ around the muon (electron) track, where $\phi$ is the azimuthal angle in radians, and is corrected for the contribution of neutral particles from PU interactions~\cite{Sirunyan:2018fpa,Khachatryan:2015hwa}.

\section{Event selection}\label{sec:event_selection}

\subsection{Trigger}\label{subsec:Trigger}

Several triggers are used to select events of interest.
One set of triggers requires \HT, the scalar sum of the \pt of all AK4 jets (with $\pt>30\GeV$ and $\abs{\eta}<2.4$) in the event, to be greater than 800, 900, or 1050\GeV, depending on the data-taking era.
In addition, single high-\pt jet triggers are used for the 2016 era.
Furthermore, events containing a high-\pt AK8 jet with trimmed mass~\cite{Krohn:2009th} in combination with either \HT calculated using AK8 jets or single AK8 jet \pt requirements are selected.
The choice of triggers is identical to the one used in a similar analysis described in Ref.~\cite{CMS:2021mjl}, where more details can be found.
The combination of all triggers (logical OR) results in a selection efficiency greater than 99\% over the full data set, for events with $\HT > 1.1\TeV$, as measured using AK8 jets in an independent sample of data events collected with a single-muon trigger and found to be consistent for both data and simulated samples.

\subsection{Preselection and signal region} \label{subsec:Preselection_and_SR}

Two stages of event selection are implemented. 
The first, the preselection, is initially applied to select events with signal-like kinematic features, comparing them to the SM backgrounds.
A more stringent selection, the signal region (SR) selection, is then applied to further improve the background rejection and define the event sample used in the statistical analysis. 
The AK8 jets will simply be referred to as ``jets'' in the following.

The preselection requires exactly three jets, corresponding to the two \PW boson candidates and the gluon candidate, for consistency all reconstructed as AK8 jets, with $\pt^{\text{j}}>200\GeV$, $\abs{\eta_\text{j}}<2.4$, and with the highest \pt jet satisfying also the $\pt^{\text{j1}}>400\GeV$ condition. 
The event \HT is required to exceed 1.1\TeV to ensure that the fully efficient trigger.
The jets with the highest and second-highest \PNetW score, designated as ja and jb, respectively, are taken as the two \PW boson candidates of the event while the third one, jc, is assumed to be the gluon. 
The probability of incorrectly assigning at least one jet in this way is approximately 20\% at the preselection level, and, \eg{}, 7\% in the SR for $\mgKK=2\TeV$ and $\MR=1\TeV$.
The masses of ja and jb, \mja and \mjb, must each be greater than 50\GeV, while no requirement is imposed on the jc mass, \mjc.
Events containing isolated leptons are rejected.

\begin{figure}[htbp]\centering
\includegraphics[width=0.45\linewidth]{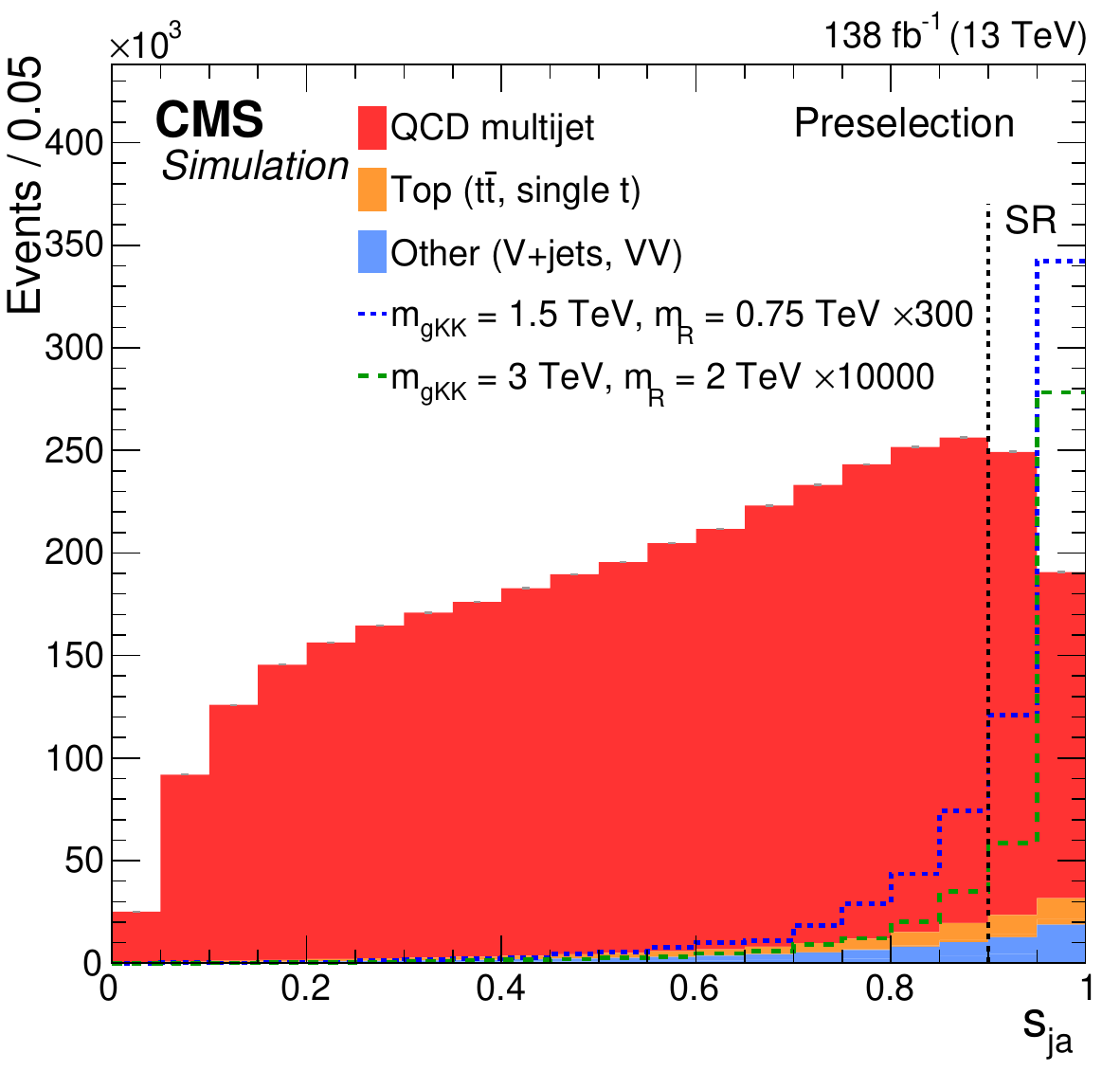}\hspace{0.05\linewidth}
\includegraphics[width=0.45\linewidth]{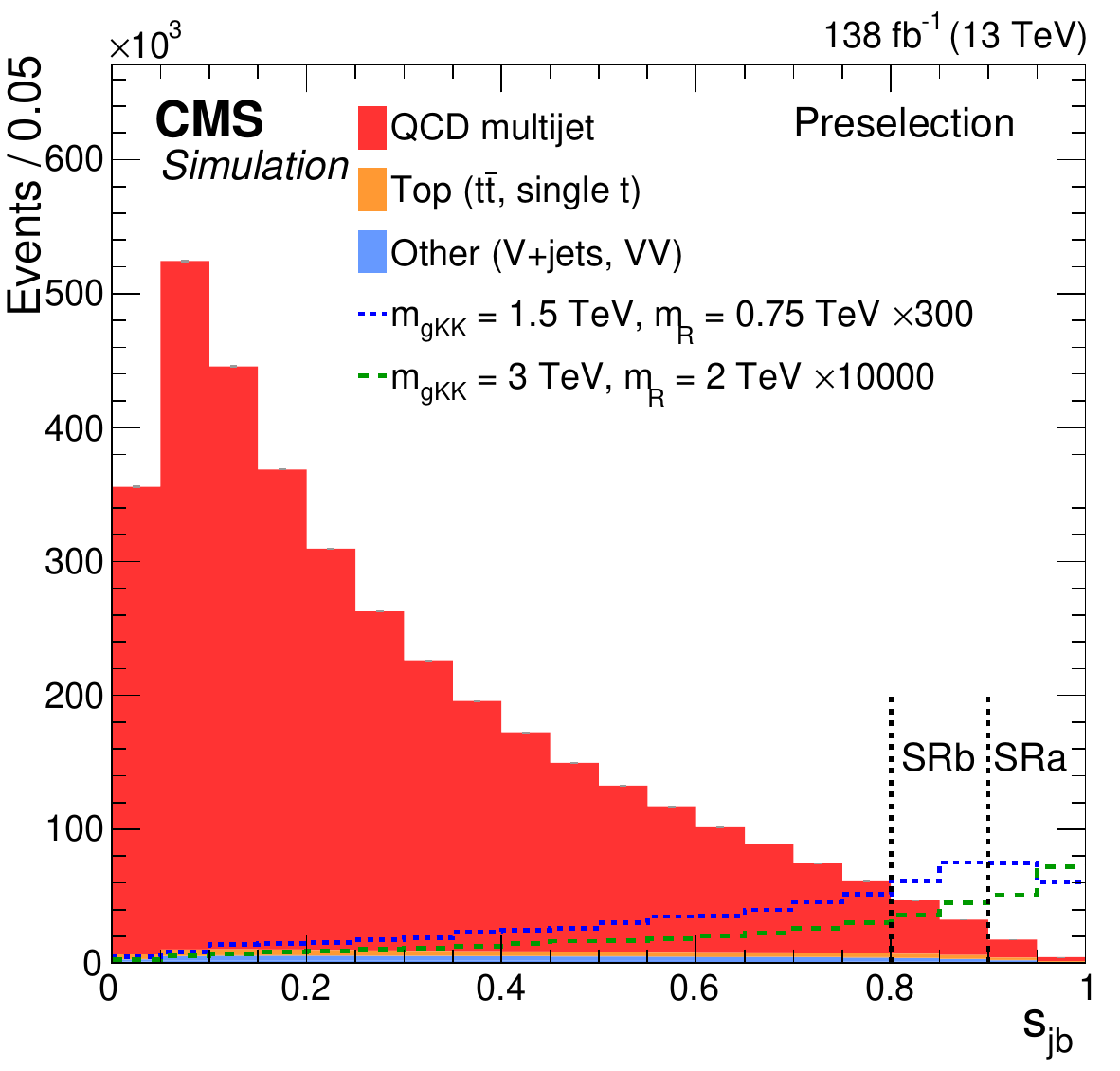}
\includegraphics[width=0.45\linewidth]{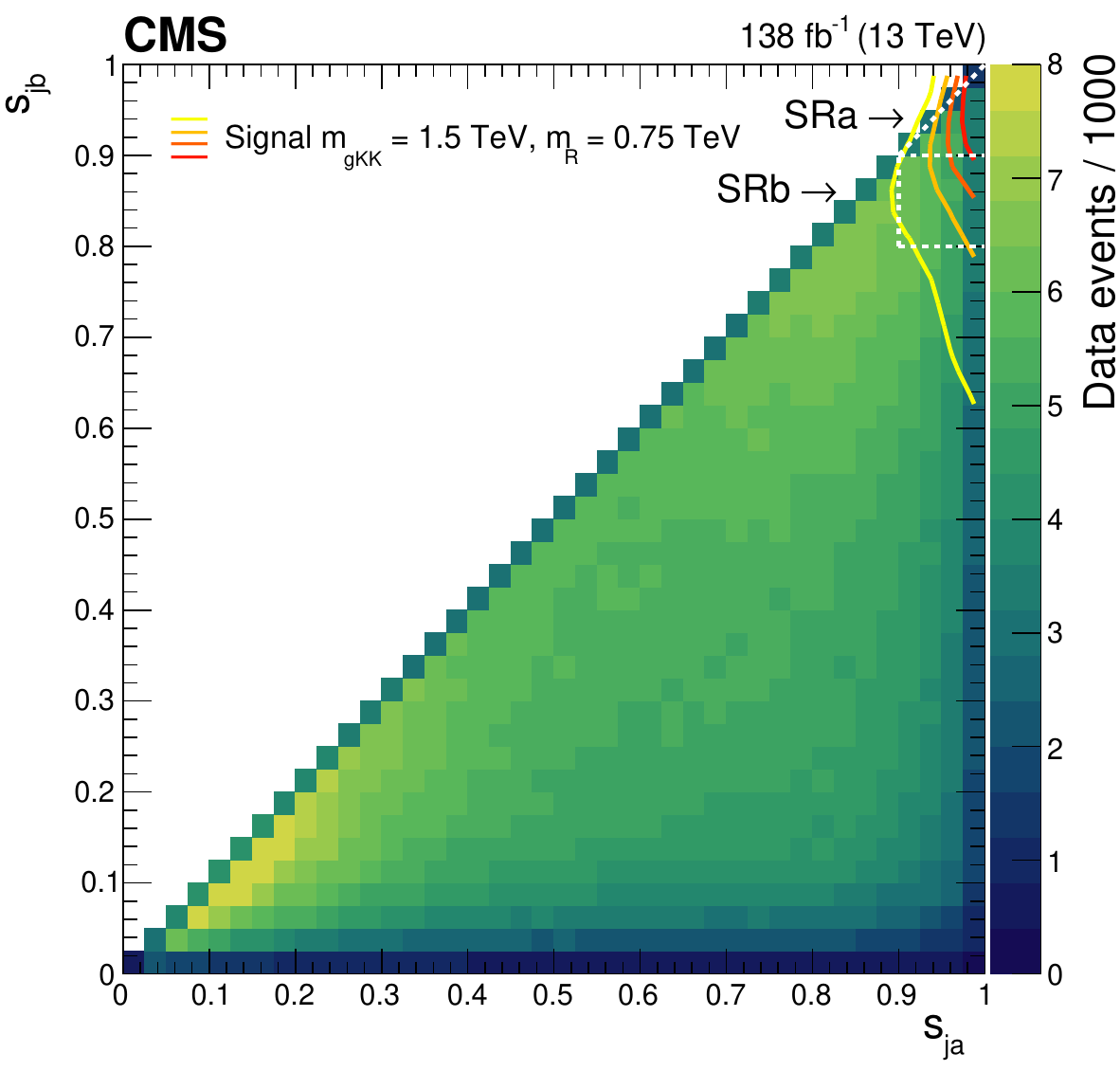}
\caption{Upper \cmsLeft (\cmsRight): the one-dimensional distributions of \sja (\sjb) for the labeled signal scenario compared to the simulated background. 
Black dashed lines indicate the SR selection boundaries.
Lower: the two-dimensional (\sja, \sjb) distribution in data. 
For the hypothesized signal ($\MR=0.75\TeV$, $\mgKK=1.5\TeV$), contours representing constant event yield fractions (80, 60, 40, and 20\% of the maximum yield) are shown to illustrate the shape.
White dashed lines indicate the boundaries between SRa, SRb samples, and regions of rejected events.
All distributions are shown after event preselection.}
\label{fig:PreselectionPlots_PNet}
\end{figure}

The ja (jb) \PNetW score, designated as \sja (\sjb), is required to be greater than 0.9 (0.8).
These conditions correspond to selection efficiencies of approximately 33 (43)\% and 0.5 (1.3)\% for signal and background jets, respectively (evaluated over a generic sample of jets other than from the analysis preselection).
The \sja and \sjb spectra for signal and background events at the preselection are presented in Fig.\ref{fig:PreselectionPlots_PNet}, upper left and right, respectively.
The combined \sja and \sjb conditions at the preselection yield a 36--45\% signal efficiency (varying with jet \pt) and 2\% (3\%) background efficiency in simulation (data).
Based on the \sjb value, events are divided into two categories, SRa and SRb, with the requirements $\sjb>0.9$ and $0.8<\sjb<0.9$, respectively. 
This binning enhances the analysis sensitivity by defining SRa with high signal purity and low background, complemented by SRb with moderate purity and higher background.

\begin{figure}[htbp]\centering
\includegraphics[width=0.412\linewidth]{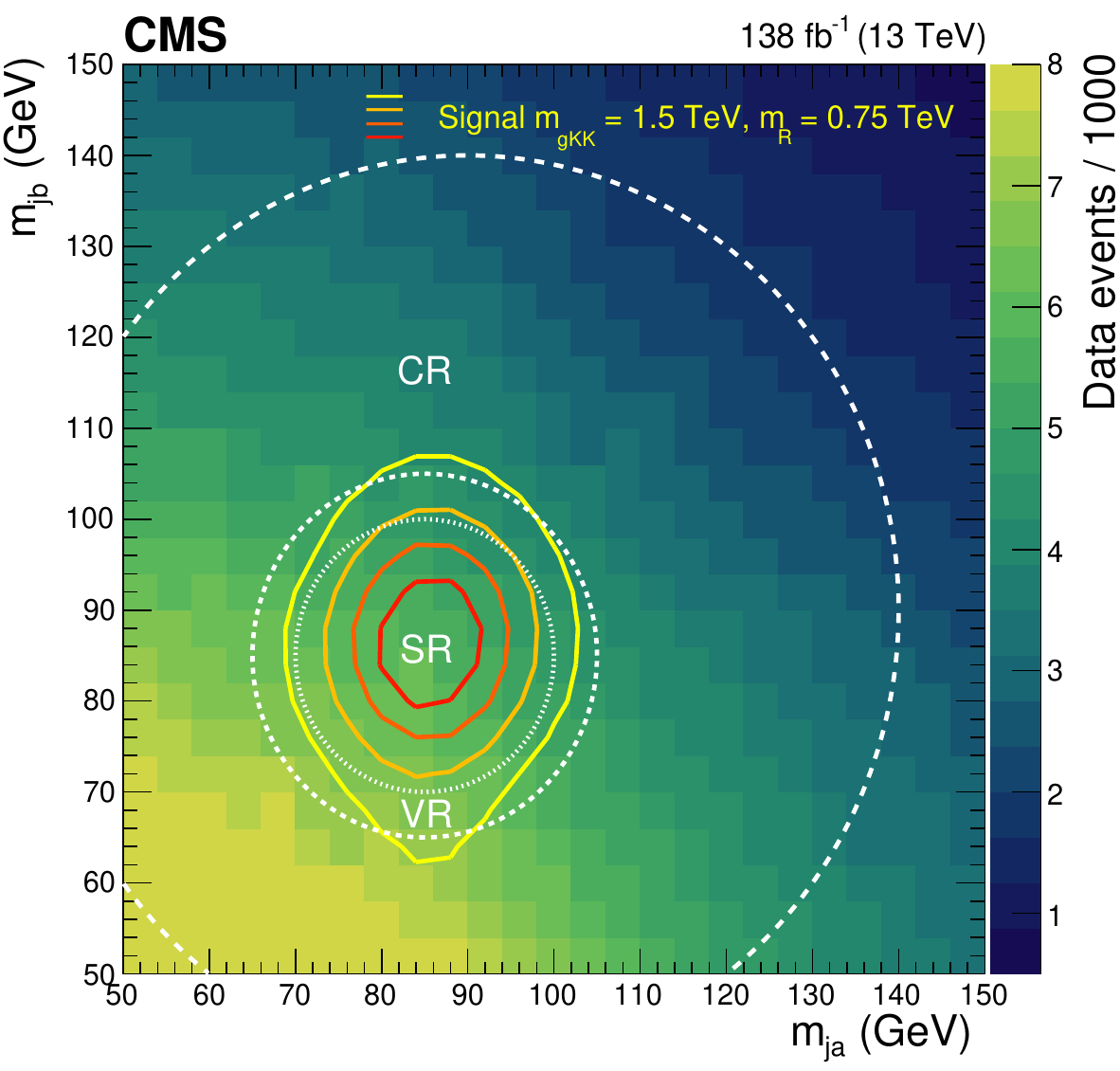}\hspace{0.05\linewidth}
\includegraphics[width=0.412\linewidth]{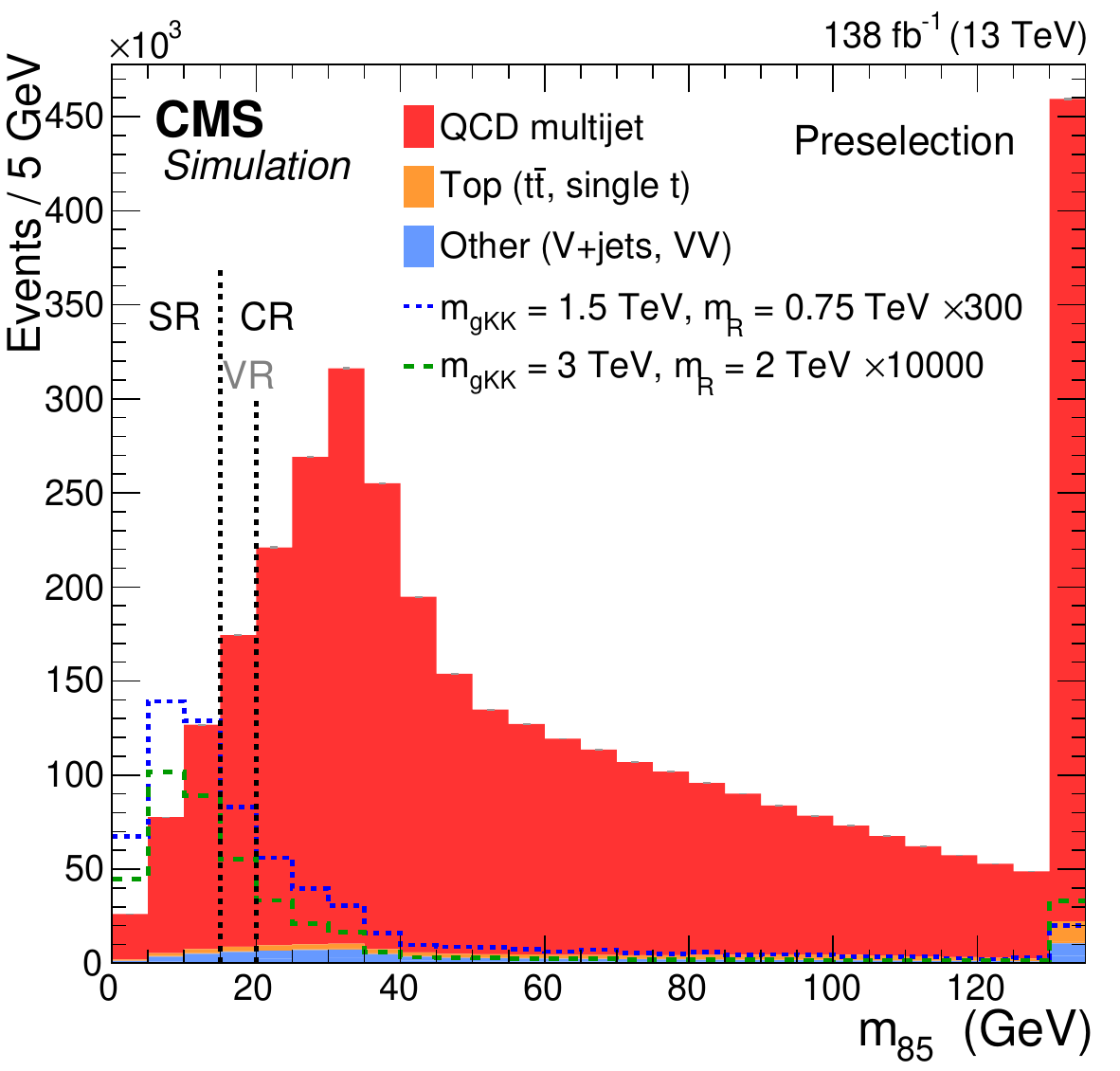}
\includegraphics[width=0.412\linewidth]{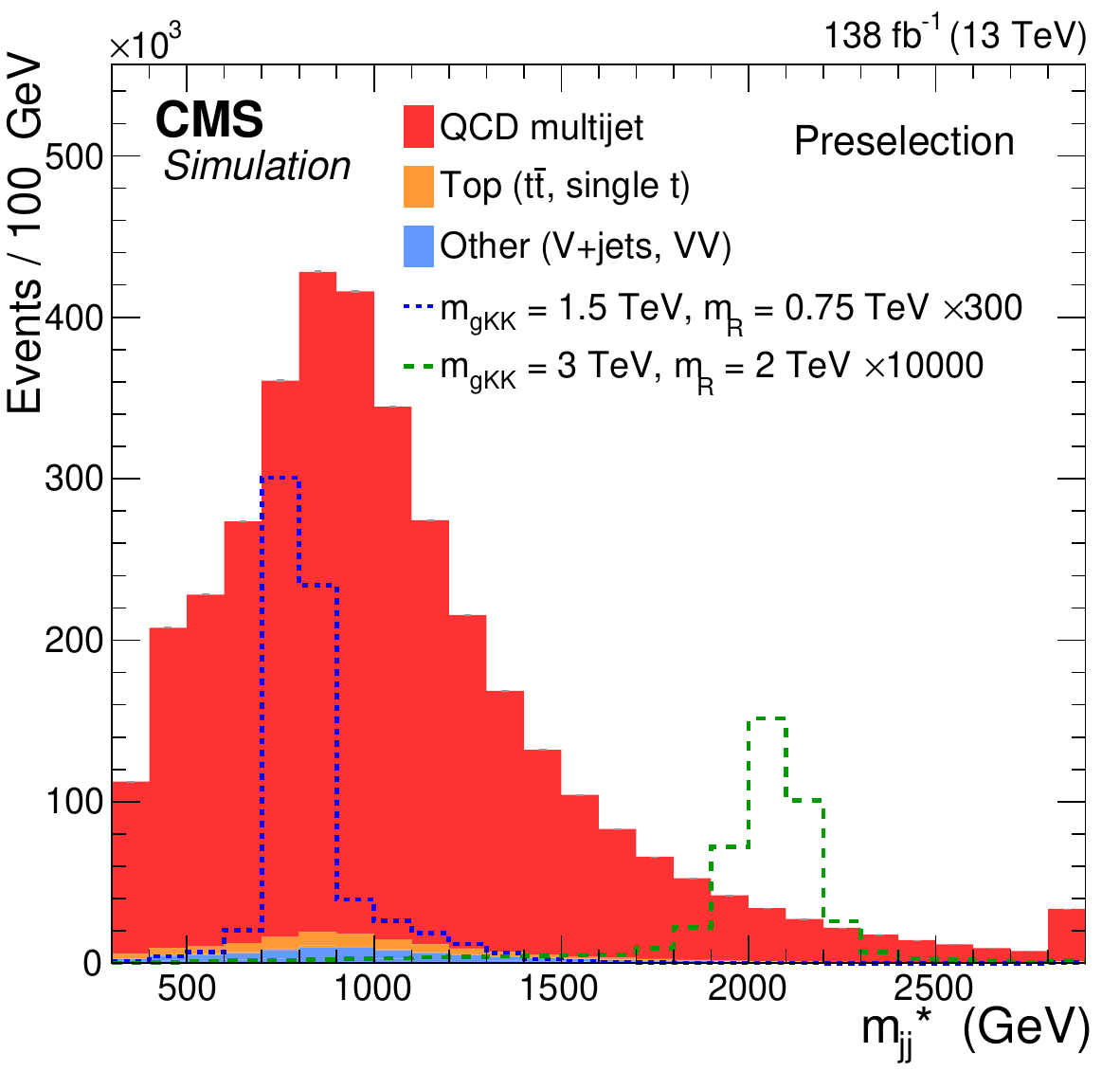}\hspace{0.05\linewidth}
\includegraphics[width=0.412\linewidth]{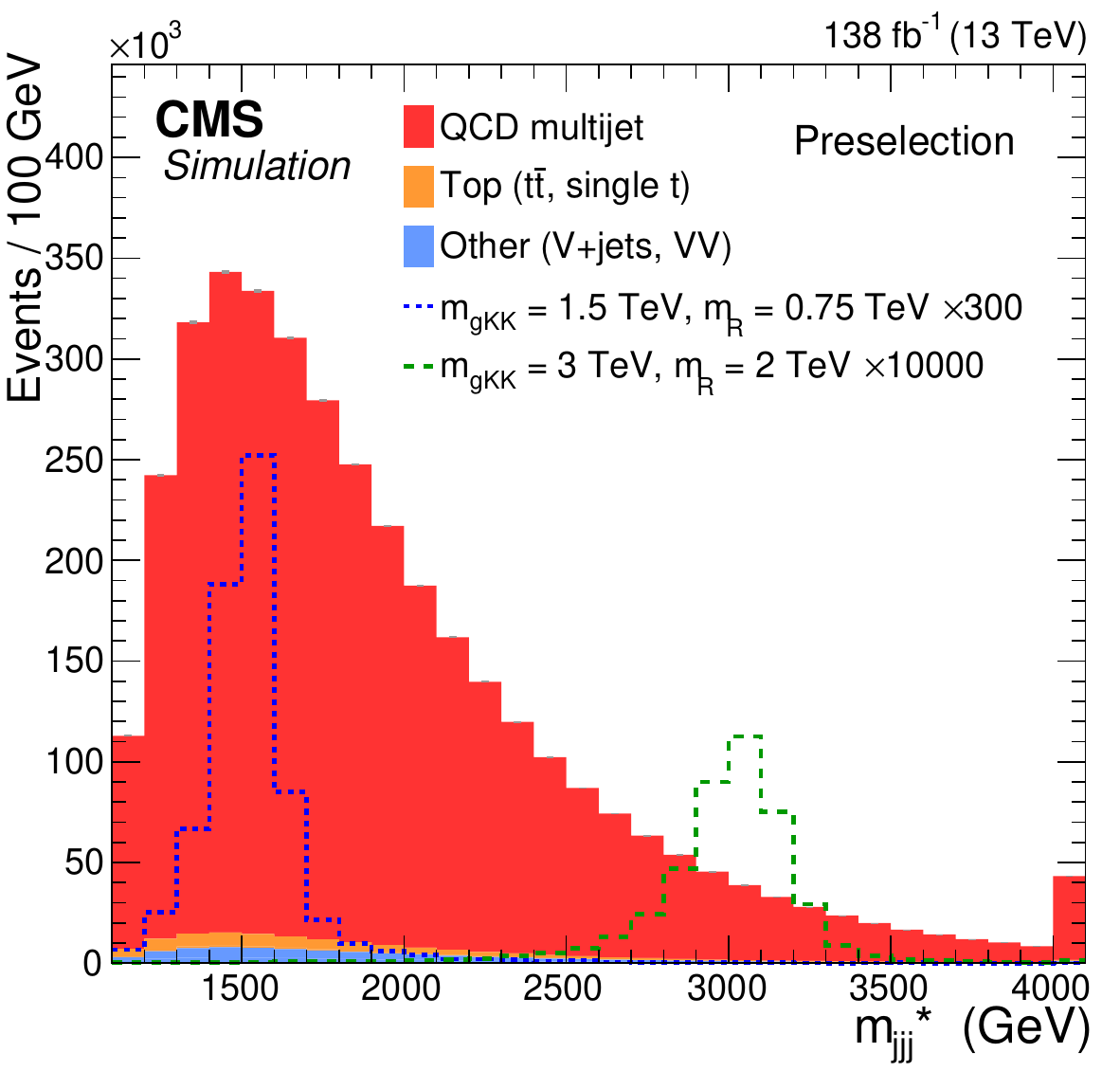}
\includegraphics[width=0.412\linewidth]{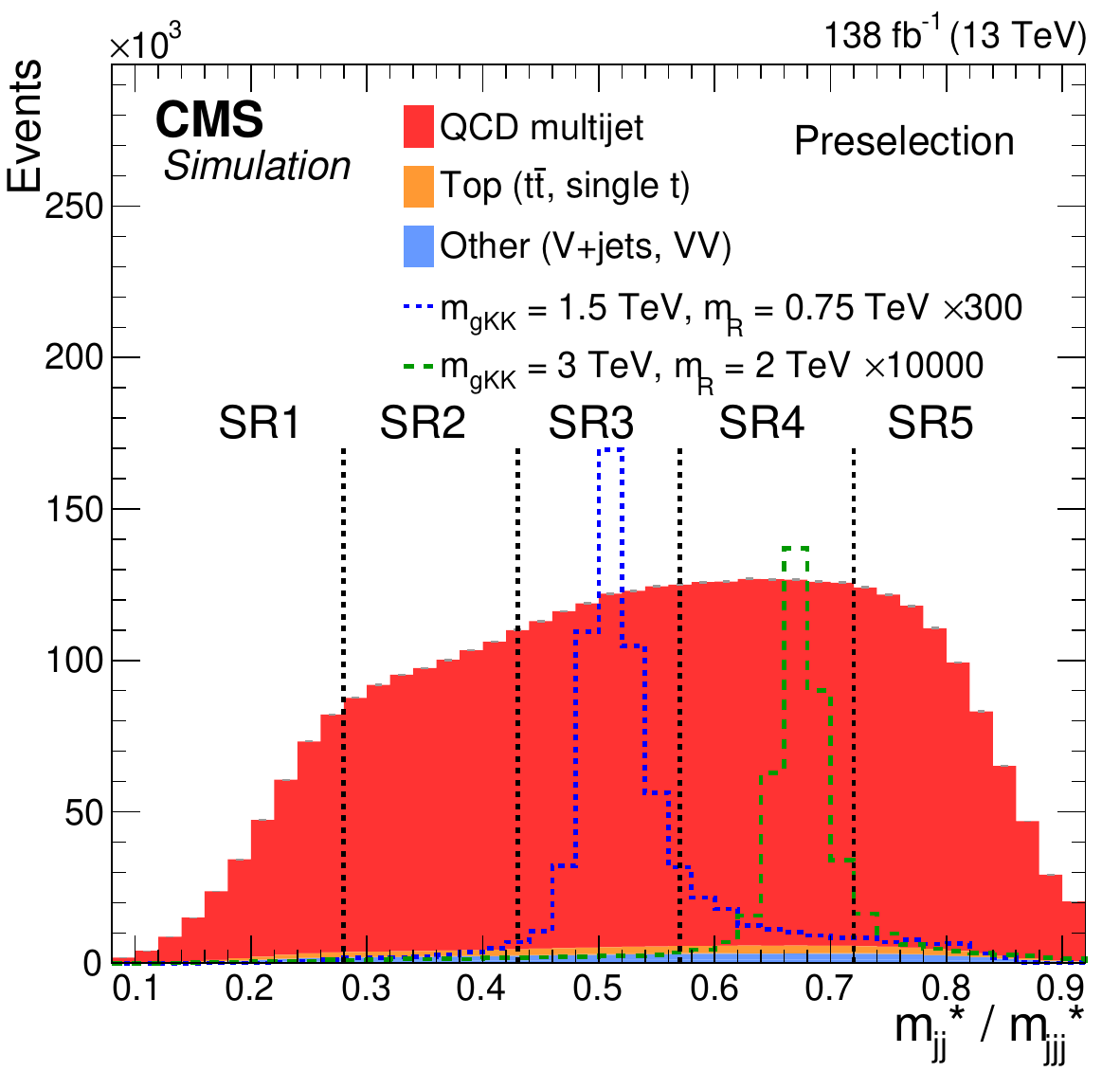}\hspace{0.05\linewidth}
\includegraphics[width=0.412\linewidth]{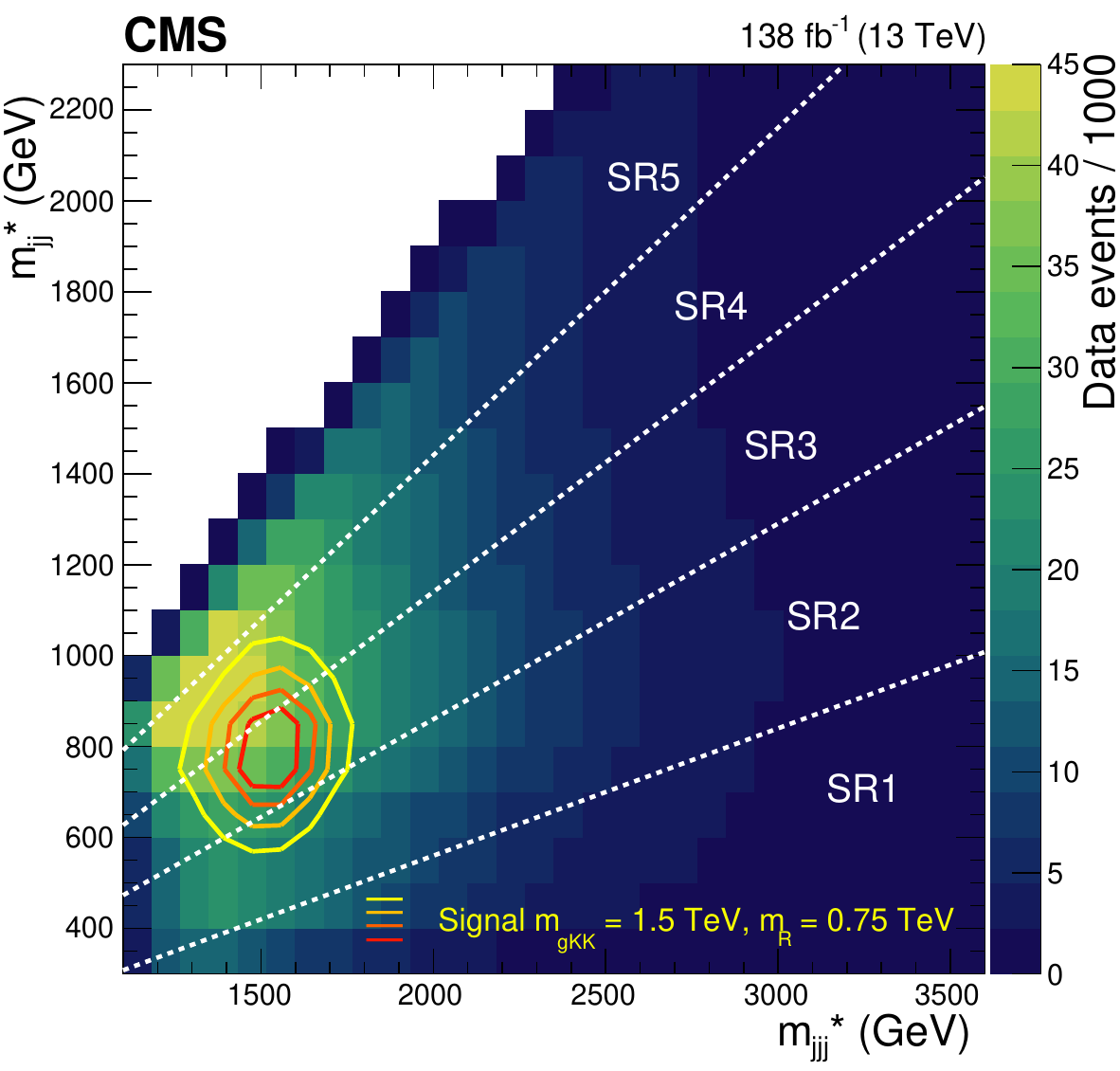}
\caption{Distributions after event preselection. Upper \cmsLeft: the two-dimensional (\mja, \mjb) distribution for a posited signal ($\MR=0.75\TeV$, $\mgKK=1.5\TeV$) compared to the background-dominated data. 
For signal, contours corresponding to 80, 60, 40, and 20\% of the maximum signal event yield are shown for shape visualization.
White dashed curves indicate boundaries between SR, VR (defined in Section \ref{subsec:QCD_prediction}), and CR areas.
The one-dimensional histograms show the \mSR (upper \cmsRight), \mjjs (middle \cmsLeft), \mjjjs (middle \cmsRight), and \mjjs/\mjjjs (lower \cmsLeft) distributions using simulated signal and background events, as indicated in the legends.
The signal events are scaled by the labeled factors times the theoretical cross sections for visibility.
Lower \cmsRight: the two-dimensional (\mjjs, \mjjjs) distribution in data. 
White dashed lines indicate the SR splitting.}
\label{fig:PreselectionPlots}
\end{figure}

We define the observable \mSR as $\mSR \equiv \sqrt{\smash[b]{(\mja-85\GeV)^2+(\mjb-85\GeV)^2}}$.
The value of 85\GeV is chosen since the reconstructed signal \PW boson mass peaks near that value, as shown in Fig.~\ref{fig:PreselectionPlots} (upper left), due to additional radiation captured within the jet cone.

We require the \PW boson jet candidates to be compatible with that mass by requiring $\mSR<15\GeV$.
Figure \ref{fig:PreselectionPlots} presents the \mja and \mjb distributions (upper \cmsLeft) and the \mSR variable  (upper \cmsRight) after the preselection.
The first and the last bins of the one-dimensional distributions of Fig.~\ref{fig:PreselectionPlots} contain the events outside the illustrated range.

The QCD multijet background is further reduced by requiring that the largest absolute difference in pseudorapidity found among the three possible jet pairs is 
$\abs{\Delta\eta_{\text{jj}}^{\text{max}}}<3$.
Events arising from \ttbar and single \PQt quark production result in final states with \PQb-tagged jets. 
To suppress this background, events that contain \PQb-tagged AK4 jets are rejected using the \textsc{DeepJet} algorithm, as described in Section~\ref{sec:Objects}.

\subsection{Resonance reconstruction and event categorization}\label{subsec:SR_binning}

The masses of the \PR and \gKK resonance candidates can be reconstructed using the invariant masses of the two \PW boson candidate jet system $m_{\text{jajb}} \equiv \mjj$, and of all three jets $m_{\text{jajbjc}} \equiv \mjjj$, respectively.
In approximately 94\% of all cases, the \PW bosons resulting from the decay of the \PR boson are correctly identified by the PNet algorithm.
For most of the remaining cases, the \PW boson decay products are not contained in a single AK8 jet.

Instead of using these two masses directly, we are using modified versions, in which the jet masses are subtracted, and the average expected \PW boson mass estimated from simulation is added.
This strategy leads to an improved signal resolution due to the detector resolution effects canceling out. 
These variables are defined and designated as $\mjjs \equiv \mjj-\mja-\mjb+2\times 85\GeV$, and $\mjjjs \equiv \Mjjj-\mja-\mjb+2\times 85\GeV$.
An improvement in sensitivity of 10 (4)\% for \mjjs (\mjjjs) is observed.
Figure \ref{fig:PreselectionPlots} (middle row) presents the corrected distributions of \mjjs (\cmsLeft) and \mjjjs (\cmsRight) in simulation after preselection.

Events are divided into five orthogonal regions based on the ratio $\mjjs/\mjjjs$ to better probe different \PR masses.
The corresponding SRs are enumerated (from lower to higher ratio values) as SR1--SR5. 
The SR1--5 ranges over \mjjs/\mjjjs are symmetrically located around 0.5, with the widths exceeding the signal resolution and dividing the data into sub-samples with approximately similar event counts, while ensuring sufficient numbers of events for QCD background predictions, as discussed in Section~\ref{subsec:QCD_prediction}.
This categorization, together with the one based on the tagger score \sjb, results in a total of ten SRs designated as SR1a, SR1b, \dots, SR5b. 
The conditions defining these SRs are summarized in Table \ref{tab:SR_Selection}.
The optimization of selection conditions, aiming to maximize the signal significance (defined as the ratio of the number of signal events to the square root of the number of background events), is evaluated for several signal scenarios at both the preselection level and in the SR, excluding one variable condition at a time.

\begin{table}[!htbp]\centering
\topcaption{Definition of the different SRs based on the \mjjs/\mjjjs ratio and \sjb.}
\begin{tabular}{l c c}
Region & \mjjs/\mjjjs & \sjb \\  \hline
SR1a & \multirow{2}{*}{$<$0.28} & $>$0.9 \\
SR1b & & 0.8--0.9 \\  [\cmsTabSkip]
SR2a & \multirow{2}{*}{0.28--0.43} & $>$0.9 \\
SR2b & & 0.8--0.9 \\  [\cmsTabSkip]
SR3a & \multirow{2}{*}{0.43--0.57} & $>$0.9 \\
SR3b & & 0.8--0.9 \\  [\cmsTabSkip]
SR4a & \multirow{2}{*}{0.57--0.72} & $>$0.9 \\
SR4b & & 0.8--0.9 \\  [\cmsTabSkip]
SR5a & \multirow{2}{*}{$>$0.72} & $>$0.9 \\
SR5b & & 0.8--0.9 \\
\end{tabular}\label{tab:SR_Selection}
\end{table}

The \mjjs/\mjjjs distribution with the SRs indicated is shown in Fig.~\ref{fig:PreselectionPlots} (lower \cmsLeft).
The two-dimensional distribution of \mjjs vs.\ \mjjjs is presented in the same figure (lower right). 
A posited signal is illustrated with constant yield contours and is centered around the generated \MR and \mgKK values.
The \mjjjs variable is used for the statistical analysis of the data and the signal extraction in these ten SRs.

\section{Background estimation} \label{sec:bkg_estimation}

The SM backgrounds in the SRs are grouped into three categories: ``QCD multijet'', ``Top'', and ``Other''.
The dominant background in all SRs stems from QCD multijet events, amounting to 55--75\% of the total yield based on simulation, depending on the SR.
Single and pair production of top quarks, collectively referred to as ``Top'', accounts for 7--17\% of total events.
All other background processes are designated as ``Other''. They make up 20--40\% of the total events, the largest contribution coming from \Wjets production. 

\subsection{Estimation of the QCD multijet background} \label{subsec:QCD_prediction}

The PNet tagger drastically suppresses the dominant background due to QCD multijet production leaving only relatively few simulated events in the SRs.
These are difficult to predict reliably using simulation.
Thus, we estimate this background contribution directly from control regions (CRs) in data.
Ten CRs are used for this purpose, each in association with a corresponding SR, \ie, CR1a--CR5b.
The selection defining the CRs is identical to that in the corresponding SRs as listed in Table~\ref{tab:SR_Selection}, except that the \mSR requirement is inverted such that $\mSR > 15\GeV$.
In addition, a new variable defined as $\mCR \equiv \sqrt{\smash[b]{ (\mja-90\GeV)^2 + (\mjb-90\GeV)^2}}$ is introduced, imposing $\mCR<50\GeV$.
In the (\mja, \mjb) plane shown in Fig. \ref{fig:PreselectionPlots} (upper left) SR and CR are indicated.
In this illustration, the area with $15<\mSR<20\GeV$, while also part of the CR, is used as a validation region (VR).
The distribution of events in the (\mja, \mjb) plane is not homogeneous. 
By using nonconcentric circles (\mSR in SRs vs. \mCR in CRs), we ensure that the average jet masses (\mja and \mjb) remain approximately the same in both the SR and CR. 
Since \mjjjs depends on \mja and \mjb, defining CRs using \mCR ensures kinematic consistency between the CR and SR, avoiding potential biases in the \mjjjs prediction.
The \mjjjs spectra in simulated QCD multijet samples are found to be consistent between SRs and CRs within their statistical uncertainties.

Based on simulation, about 90\% of the events in the CRs are QCD multijet events.
Their kinematic properties are very similar to those of the QCD multijet events in the SRs.
Most importantly, CR events feature the same selection efficiency as the SR events, as the \PNetW tagger condition remains identical.

The \mjjjs spectra of QCD multijet events in SRs are predicted using the data in the CRs.
All other SM processes labeled as ``Rest'' in the following are subtracted using simulation. Each resulting CR template is scaled to the event yields of the corresponding SR using the ratio of total simulated QCD multijet events in the given SR and CR, $\text{QCD}_{\text{SRxy}}/\text{QCD}_{\text{CRxy}}$. 
This can be written as:
\begin{equation}\label{eq:Pred} 
\text{Pred}_{\text{SRxy}}^{\text{QCD}}\ \equiv \ [\text{Data}-\text{Rest}]_{\text{CRxy}}\frac{\text{QCD}_{\text{SRxy}}}{\text{QCD}_{\text{CRxy}}}.
\end{equation}
The QCD multijet event yields obtained in this way are factors of 1.6--1.8 larger than predicted by the simulation depending on the SR. These differences are consistent with similar analyses \cite{CMS:2022lqh, CMS:2021mjl}, reflecting the limitations of QCD multijet modeling at high tagger scores.

\subsection{Estimation of the Top and Other backgrounds} \label{subsec:top_other_prediction}

The normalization and the shape of the Top background is estimated from simulation and validated in data CRs.
For this purpose, we are using the data from ten regions, defined similarly to the SRs, but requiring the presence of at least one \PQb-tagged AK4 jet. These \PQb-tagged jet enriched regions are called bRs in the following.
The signal contribution in the bRs is an order of magnitude lower than in the SRs, and thus negligible.
According to simulation, the bRs consist to approximately 70, 25, and 5\% of Top, QCD multijet, and Other processes, respectively.
The QCD multijet contribution in each bR$_{xy}$ is estimated using simulation, corrected by the factor $[\text{Data} - \text{Rest}]_{\text{CRxy}} / \text{QCD}_{\text{CRxy}}$ obtained from the corresponding CRxy.
The remaining difference with respect to the data in each bR, ranging from 0 to 50\%, is then attributed to the Top background.
A 50\% rate uncertainty in the Top production cross section, uncorrelated across the SRs, is assigned.
The Top background \mjjjs shapes in the bRs are found to be consistent with simulation within uncertainties.

The contribution of the Other background in the SRs is estimated from simulation. 
It is dominated by \Wjets production, which accounts for approximately two thirds of the events.
The selections used in this analysis, in particular the vetoing of leptons and selecting two \PW boson candidates, do not allow the definition of a pure \Wjets sample close to the SR. 
However, a similar triboson analysis~\cite{CMS:2022lqh}, for which \Wjets production was the dominant background, found this process to be well-described by simulation. 
A conservative 30\% uncertainty is assigned to the Other background rate.
A correction is applied to both Top and Other background simulations to account for the differences in the PNet \PW boson selection efficiency in data and simulation, as described below.

\subsection{Calibration of the \texorpdfstring{\PW}{W} boson tagger} \label{subsec:calibration}

For the QCD multijet background, an estimate using the same selection conditions on \sja and \sjb 
in the data CRs and SRs, obviates the need for correcting the PNet selection efficiency.
For the signal and all other background processes that contain genuine \PW bosons, the PNet selection efficiency is calibrated using the data.
By requiring the data to match the simulation in the jet mass distribution, scale factors (SFs) are derived 
for the selected \PW boson candidate jet with \PNetW scores of 0.8--0.9 and $>$0.9.
These SFs are binned in jet \pt and summarized in Table~\ref{tab:SFs}.

\begin{table}[ht]\centering
\topcaption{The \PNetW scale factors calculated for exclusive \pt and \PNetW tagger score ranges of \PW boson matched jets.}
\begin{tabular}{l c c c}
Jet \pt ({\GeVns}) & 200--300 & 300--400 & $>$400  \\
\hline
$\sja>0.9$     & $0.83 \pm 0.03$ & $0.84 \pm 0.04$ & $0.82 \pm 0.05$ \\
$0.8<\sjb<0.9$ & $1.08 \pm 0.03$ & $1.01 \pm 0.04$ & $1.02 \pm 0.05$ \\
\end{tabular}
\label{tab:SFs}
\end{table}

To apply the SFs, all reconstructed jets in simulated events are subjected to matching with the generated partons using geometrical conditions, as described in Ref.~\cite{CMS:2021mjl}. 
Both the simulated signal and background events are corrected with these SFs for each jet matched as a \PW boson, across all samples used in the analysis and in all regions.

\subsection{Background estimation validation} \label{subsec:Validation}

The background estimation method is validated in the data without using the actual SR events.
Instead, SRs are replaced by ten VRs. 
These are obtained using the SR selection conditions given in Table~\ref{tab:SR_Selection}, changing only the $\mSR<15\GeV$ condition to $15<\mSR<20\GeV$, as illustrated in Fig.~\ref{fig:PreselectionPlots}, upper row.
As the $15<\mSR<20\GeV$ range nominally is part of the CRs, for this test only, the CRs are redefined excluding these overlapping events. 
The prediction of the shape and rate of QCD multijet production is evaluated for each VR with the same method as used for the SRs.
Simulation is used for processes other than QCD multijet, as their contribution is small outside the \PW boson mass window.
A maximum likelihood fit is performed, including all the sources of systematic uncertainty described in Section~\ref{sec:uncertainties} for the actual result interpretation.
The observed \mjjjs distributions in data agree with the predicted ones within statistical uncertainties, validating the prediction method described in Section~\ref{subsec:QCD_prediction}.
The maximum nonclosure observed in the VRs prior to the fit is assigned as the uncertainty in the QCD background rate.

\section{Systematic uncertainties} \label{sec:uncertainties}

Systematic uncertainties are taken into account for the background estimation and the signal extraction.
For each source of uncertainty, a nuisance parameter is assigned, which is constrained by the data.
These are summarized in Table~\ref{tab:SysUnc}.
Uncertainties in the background estimation and signal modeling are grouped separately.
It is also indicated whether the uncertainty is in the \mjjjs shape or the rate of a particular process.

\begin{table}[!htbp]\centering
\topcaption{The sources of systematic uncertainties accounted for in the analysis. 
From left to right: indication of whether an uncertainty is evaluated for the background (B) or signal (S), whether the uncertainty affects shape or rate, magnitudes (where applicable, or standard deviations ($\sigma$)), and the total number of nuisance parameters (NPs) used along with their correlations across SRs.
The ``*'' indicates a value or a shape template different for each signal scenario.}
\cmsTable{
\begin{tabular}{l c c c c}
Uncertainty source & B or S & Effect on & Magnitude & Number of NPs \& correlations  \\
\hline
Normalization QCD multijet  & B & Rate & 20\% & 10, uncorr.\ across SRs\\
Normalization Top  & B & Rate & 50\% & 10, uncorr.\ across SRs\\
Normalization Other & B & Rate & 30\% & 10, uncorr.\ across SRs\\ [\cmsTabSkip]
QCD multijet B shape due to \mCR usage & B & Shape & $\pm 1\sigma$ templates & 10, uncorr.\ across SRs\\
QCD multijet B shape due to other processes & B & Shape & $\pm 1\sigma$ templates & 10, uncorr.\ across SRs\\ [\cmsTabSkip]
PU reweighting \& integrated luminosity & S & Rate & 1.7\% & 1, correlated across all SRs\\
PDFs & S & Rate & $<$10\% & 1, correlated across all SRs *\\
$\mu_\text{R}$ and $\mu_\text{F}$ scales & S & Rate & $<$0.8\% & 1, correlated across all SRs *\\
\PNetW selection efficiency per jet (event) & S & Shape & 6\% (12\%) & 1, correlated across all SRs\\ [\cmsTabSkip]
JES & S & Shape & $\pm 1\sigma$ templates & 1, correlated across all SRs *\\
JER & S & Shape & $\pm 1\sigma$ templates & 1, correlated across all SRs *\\
\end{tabular}
}
\label{tab:SysUnc}
\end{table}

\subsection{Systematic uncertainties in the background estimation} \label{sec:uncertainties_for_BKG}

\subsubsection{Systematic uncertainties in the background rates}

The three background categories (defined as QCD multijet, Top, and
Other) are assigned normalization rate uncertainties of magnitude 20, 50, and 30\%, respectively, with a log-normal prior.
One nuisance is used for each background category and for each SR, resulting in a total of 30 nuisance parameters.
The 20\% magnitude of the QCD multijet background category is obtained from the maximum nonclosure observed in the VRs before performing the fit and from the statistical uncertainty in the ratio factor of Eq.~(\ref{eq:Pred}), resulting from simulation. 
The observed modeling of the \PNetW score and the \mjjs/\mjjjs distributions at the preselection level supports the choice of using uncorrelated nuisance parameters across the SRs.
A 20\% uncertainty is assigned to account for residual differences between the simulation and the data.
The 50\% magnitude of the Top category is assessed by the maximum nonclosure observed in the bRs. 
The 30\% magnitude of the Other category, since it is dominated by the \Wjets production, is assessed from a similar analysis~\cite{CMS:2022lqh} where the \Wjets process is the dominant background.
Uncertainties in the JES and JER of the simulated backgrounds were found to have a negligible impact on the final result.

\subsubsection{Systematic uncertainties in the QCD multijet background shape}

The \mjjjs shape of the QCD multijet background is assigned two uncertainties.
As the shape comes effectively from the data in the CRs, as Eq.~(\ref{eq:Pred}) denotes, the prediction is susceptible to potential biases due to:
\begin{enumerate}
\item[(a)] the choice of the circular area in the (\mja, \mjb) plane used to define the CRs, which is centered around (90, 90)\GeV (Fig.~\ref{fig:PreselectionPlots} upper left), \ie{}, the choice of using \mCR, and
\item[(b)] the subtraction of the largely unknown rate of backgrounds other than QCD multijet production in the prediction method (given by Eq.~(\ref{eq:Pred})). 
\end{enumerate}
For (a), we derive two alternative shape predictions ($\pm 1\sigma$ variations) using two modified CR selection conditions: $\mSR<50\GeV$, and $m_{95}<50\GeV$, where $m_{95}$ is defined using 95\GeV in analogy with \mSR. 
In this way, we parametrize the CR definition based on the center of CR area allowing the prediction to be only loosely dependent on the nominal \mCR used.
For (b), to account for a potentially inaccurate normalization of the Other processes in CR, we derive the QCD multijet shape estimate scaling the Rest events in Eq.~(\ref{eq:Pred}) by factors of zero and two, also obtaining up and down variations of the \mjjjs shape. 

A nuisance parameter with a Gaussian prior is assigned for each of the two sources of uncertainty and for each of the ten SRs, resulting in 20 nuisances.
These QCD multijet shape uncertainties provide flexibility in particular for the tails of the \mjjjs spectra to adapt to a potentially softer or harder spectrum observed in the SR.

\subsubsection{Systematic uncertainties in the signal rate}

The simulated PU distribution is scaled to match the data using an effective total inelastic cross section of 69.2\unit{mb}.
The uncertainty in this procedure is evaluated by varying the total inelastic cross section by $\pm$4.6\%~\cite{Sirunyan:2018nqx}.
This results in a 0.5\% uncertainty in the signal normalization in the SRs with negligible shape effects.
The uncertainty in the integrated luminosity individually for the 2016, 2017, and 2018 data-taking years amounts to 1.2--2.5\%~\cite{Sirunyan:2021qkt,CMS-PAS-LUM-17-004,CMS-PAS-LUM-18-002}.
The overall uncertainty for the 2016--2018 period is 1.6\%.
The two sources are combined to a total uncertainty of 1.7\%, implemented with a log-normal prior.

The renormalization $\mu_\text{R}$ and factorization $\mu_\text{F}$ scales, and PDF uncertainties affecting the signal selection efficiency are evaluated per SR and mass point.
The scale uncertainties are obtained by varying $\mu_\text{R}$ and $\mu_\text{F}$ independently by factors of 1/2 and 2 (without considering the extreme cases of the opposite-direction variations).
Half of the maximum difference of these variations is taken as the uncertainty with a log-normal prior, resulting in a $<$0.8\% effect on the signal rate.
The PDF uncertainties are evaluated using the replicas of the NNPDF PDF set, for each signal scenario individually. 
The corresponding Hessian members of the PDF set are varied, and the quadratic sum of the variations from the central value is calculated for each SR.
The resulting variation in signal event yields is 0--10\% for different \gKK masses.
For the overall signal scale uncertainty, a single nuisance parameter is used to account for both the $\mu_\text{R}$ and $\mu_\text{F}$ scales, and PDF uncertainties with a log-normal prior.

\subsubsection{Systematic uncertainties in the signal shape}

The JES uncertainty is evaluated by collectively varying the reconstructed four-momenta of all jets in an event, according to their $\eta$- and \pt-dependent associated uncertainties. 
The resulting impact on the \mjjjs distributions is then assigned as the uncertainty.
Moreover, to properly evaluate the systematic effect coming from differences in the JER between data and simulation, smearing is also applied to the latter by varying the jet resolutions according to their uncertainties.
Both uncertainty sources are implemented as shape-altering nuisance parameters using Gaussian priors.

\begin{figure}[!htbp]\centering
\includegraphics[width=0.45\linewidth]{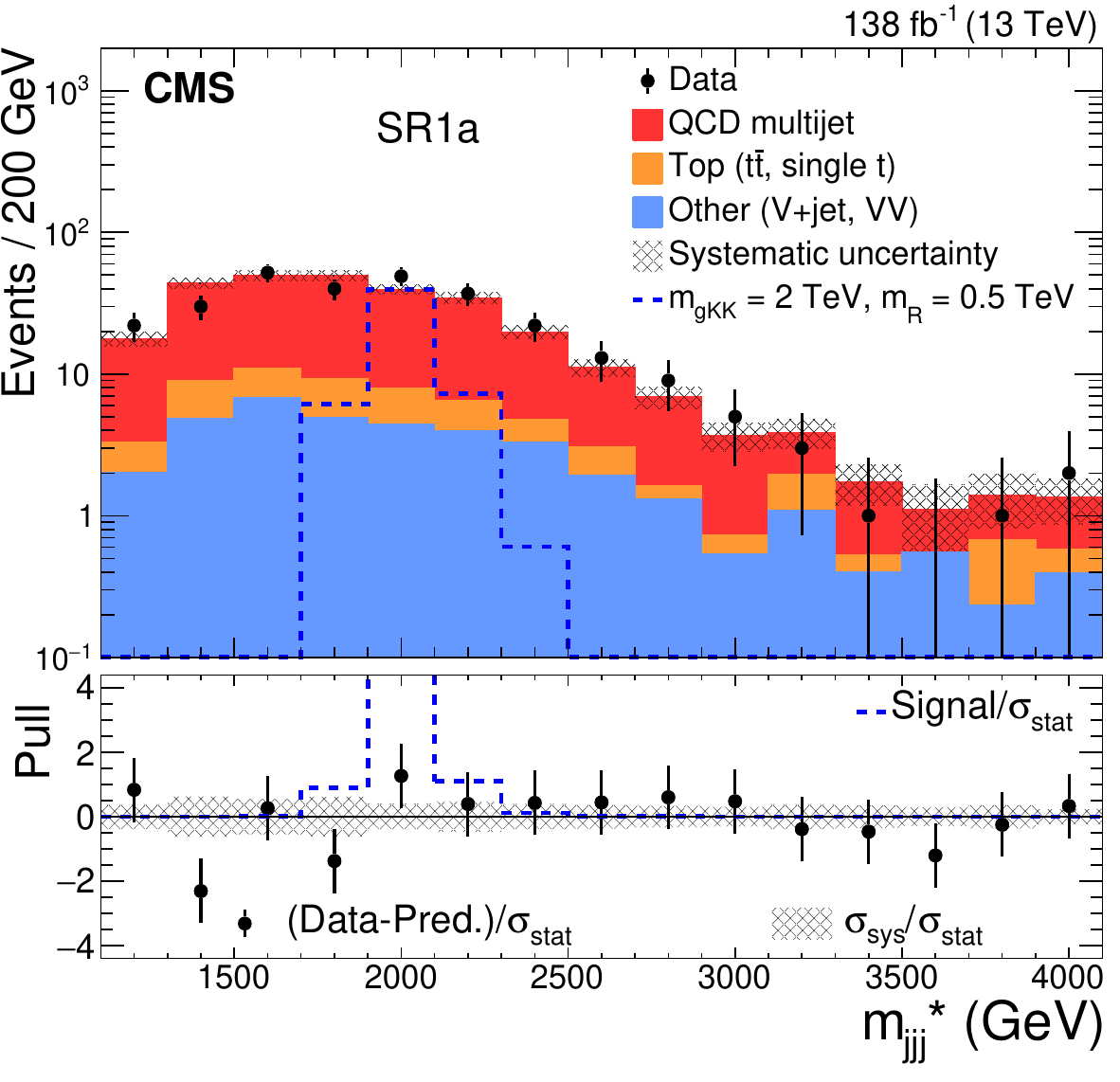}
\includegraphics[width=0.45\linewidth]{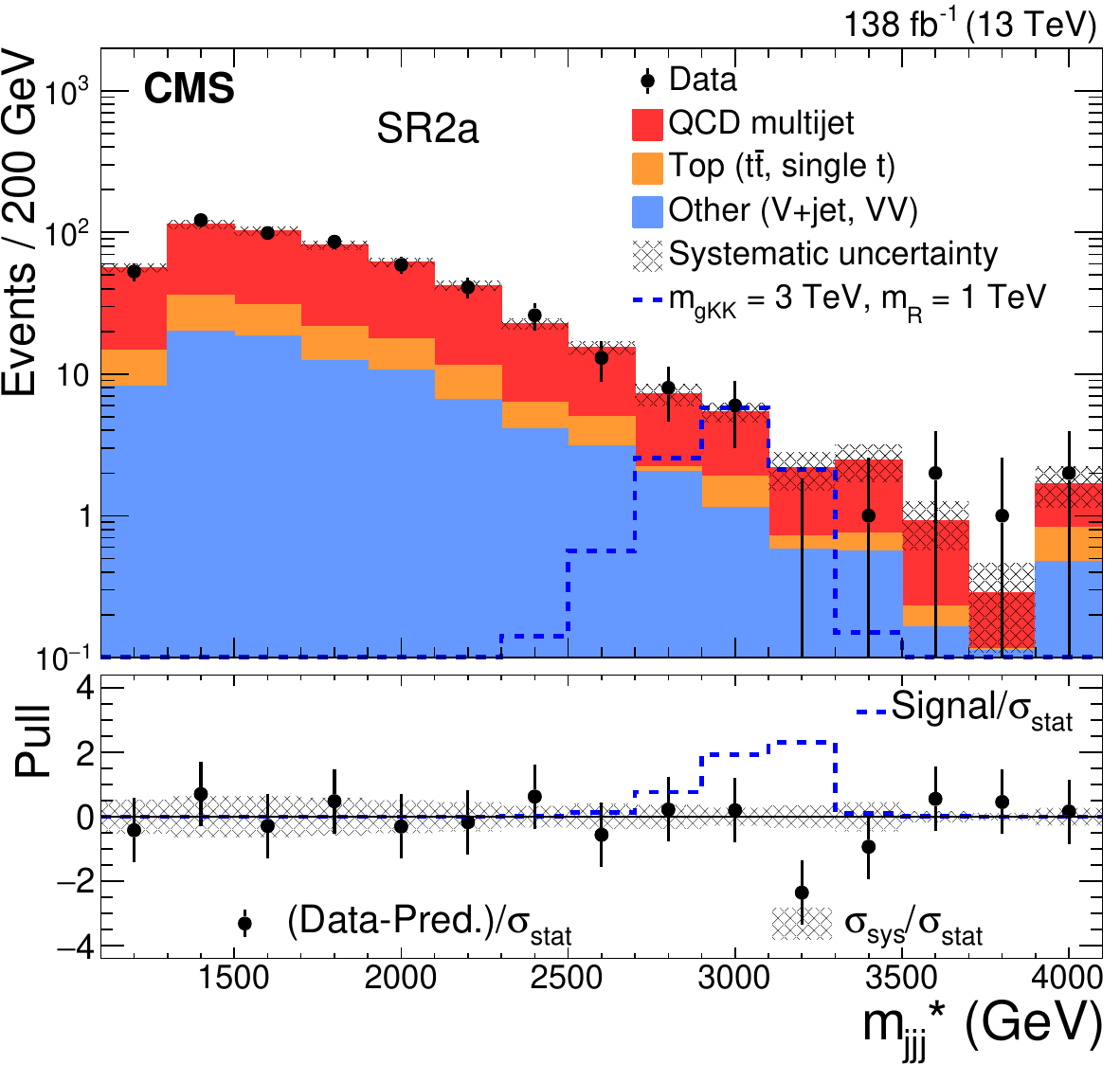}
\includegraphics[width=0.45\linewidth]{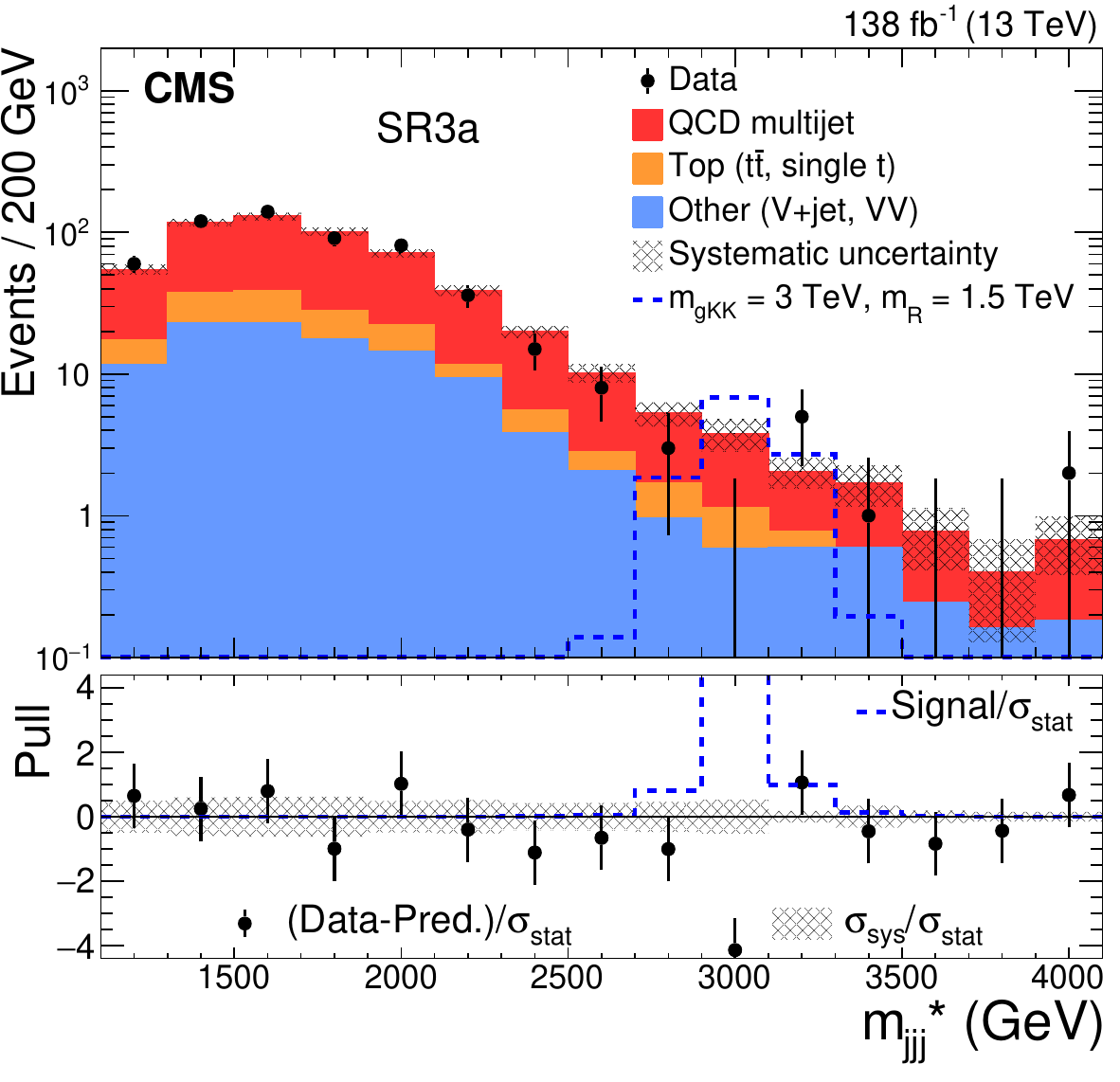}
\includegraphics[width=0.45\linewidth]{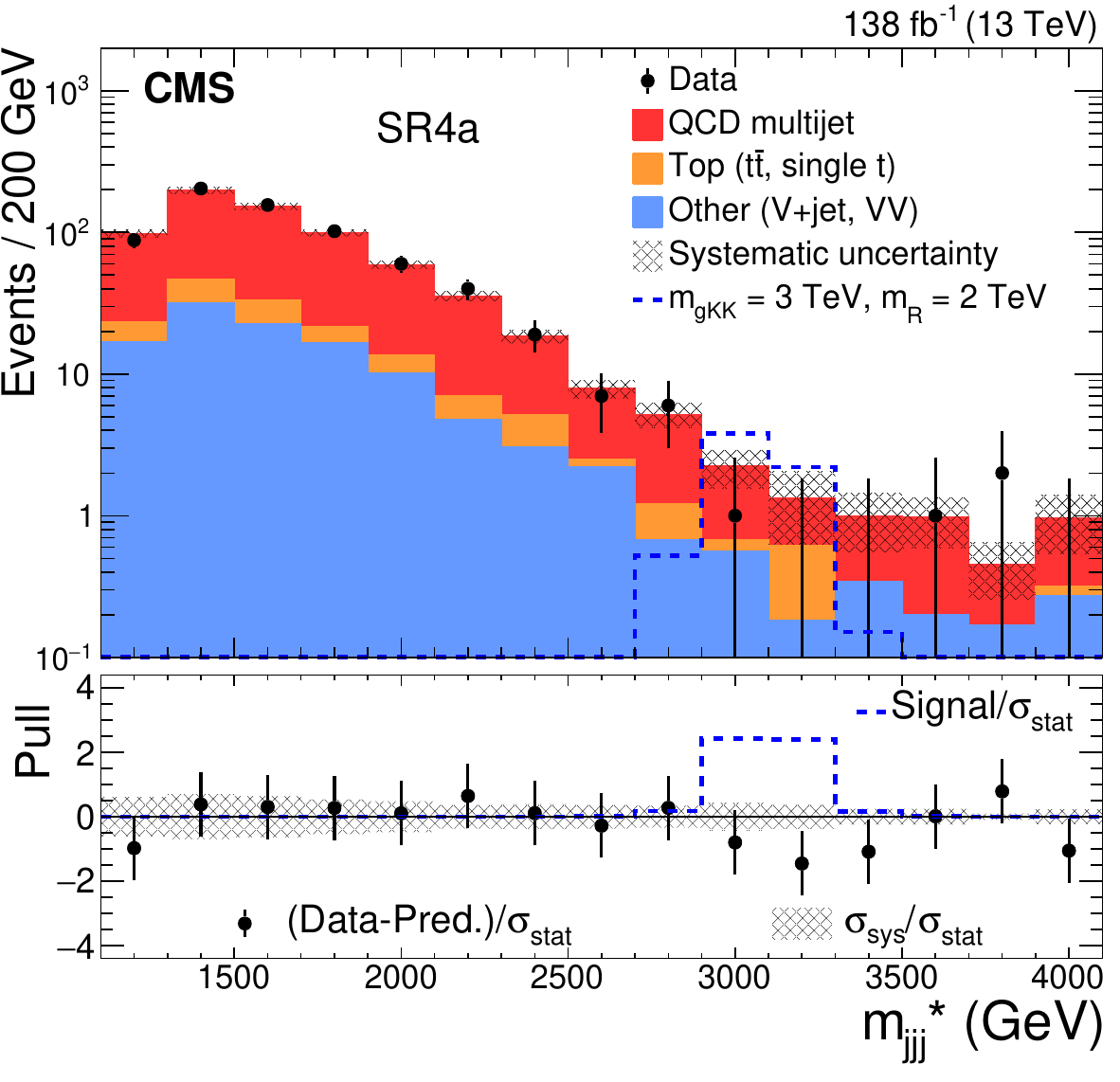}
\includegraphics[width=0.45\linewidth]{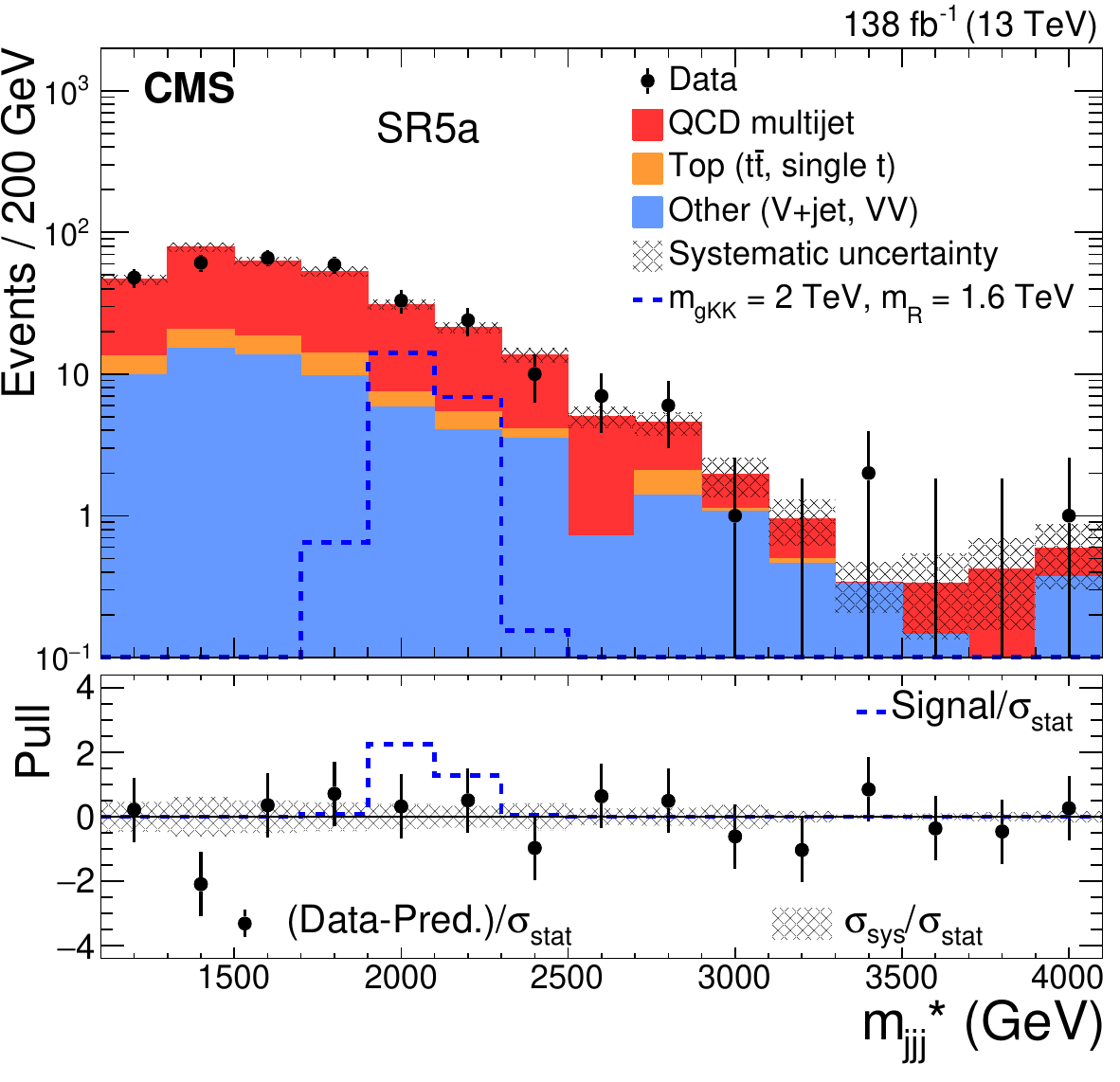}
\caption{The \mjjjs postfit spectra in the five SRxa.
Upper to lower, and left to right: SR1a, SR2a, SR3a, SR4a, and SR5a.
The lower panels show the pull distributions, as defined in the text, together with expected signal event yields and the systematic uncertainty normalized to $\sigma_{\text{stat}}$.}
\label{fig:Postfit_a}
\end{figure}

\begin{figure}[!htbp]\centering
\includegraphics[width=0.45\linewidth]{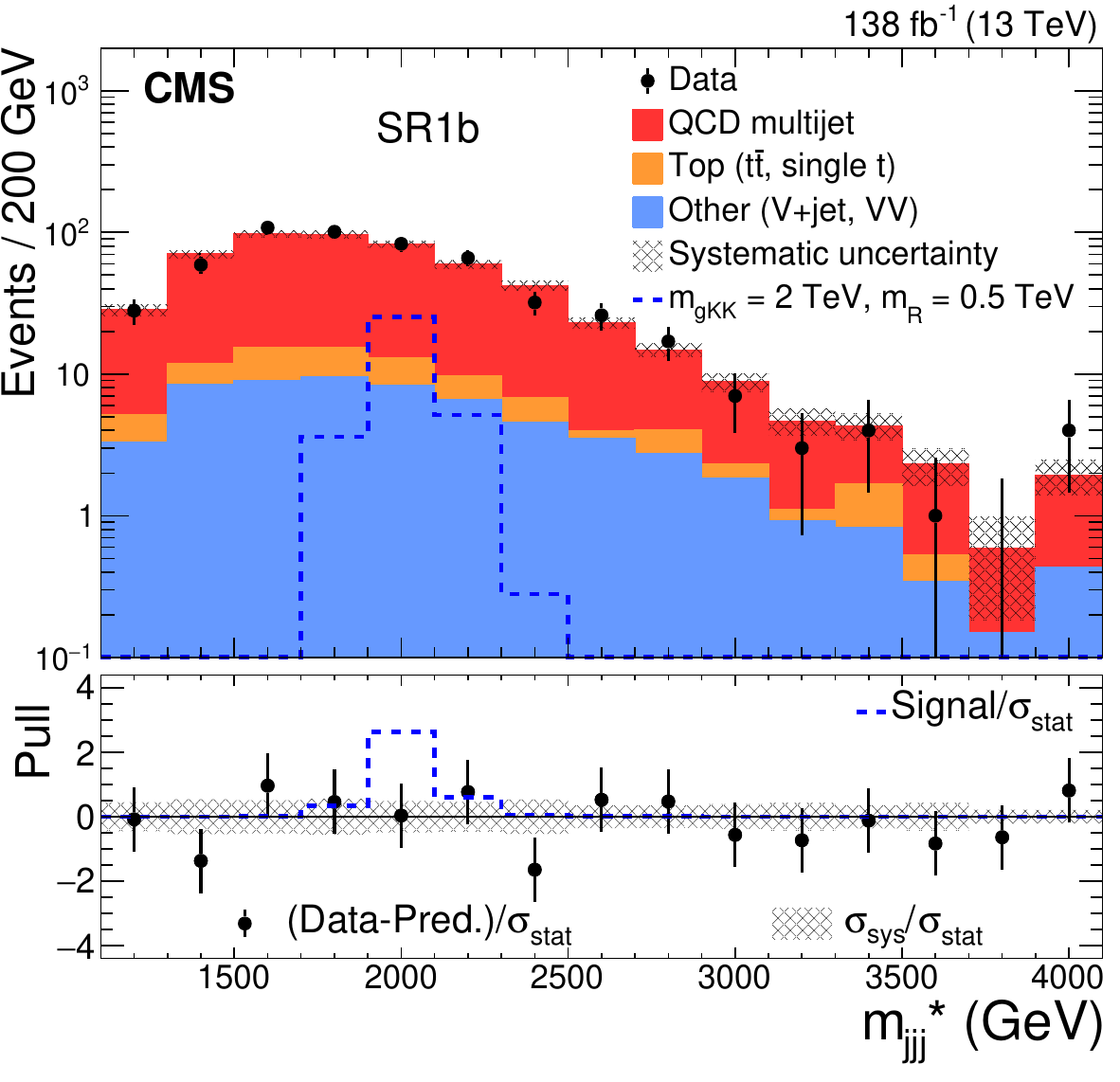}
\includegraphics[width=0.45\linewidth]{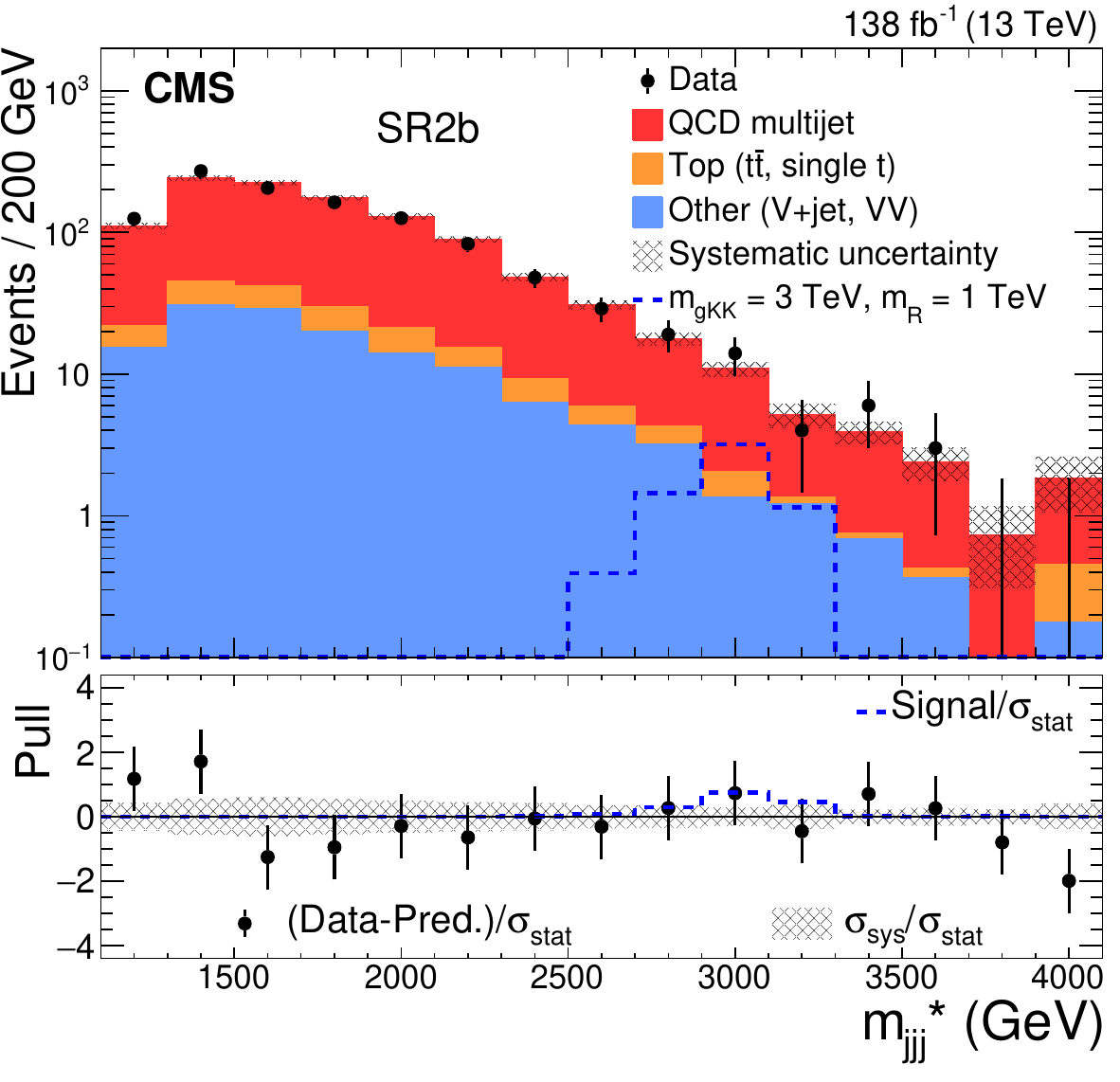}
\includegraphics[width=0.45\linewidth]{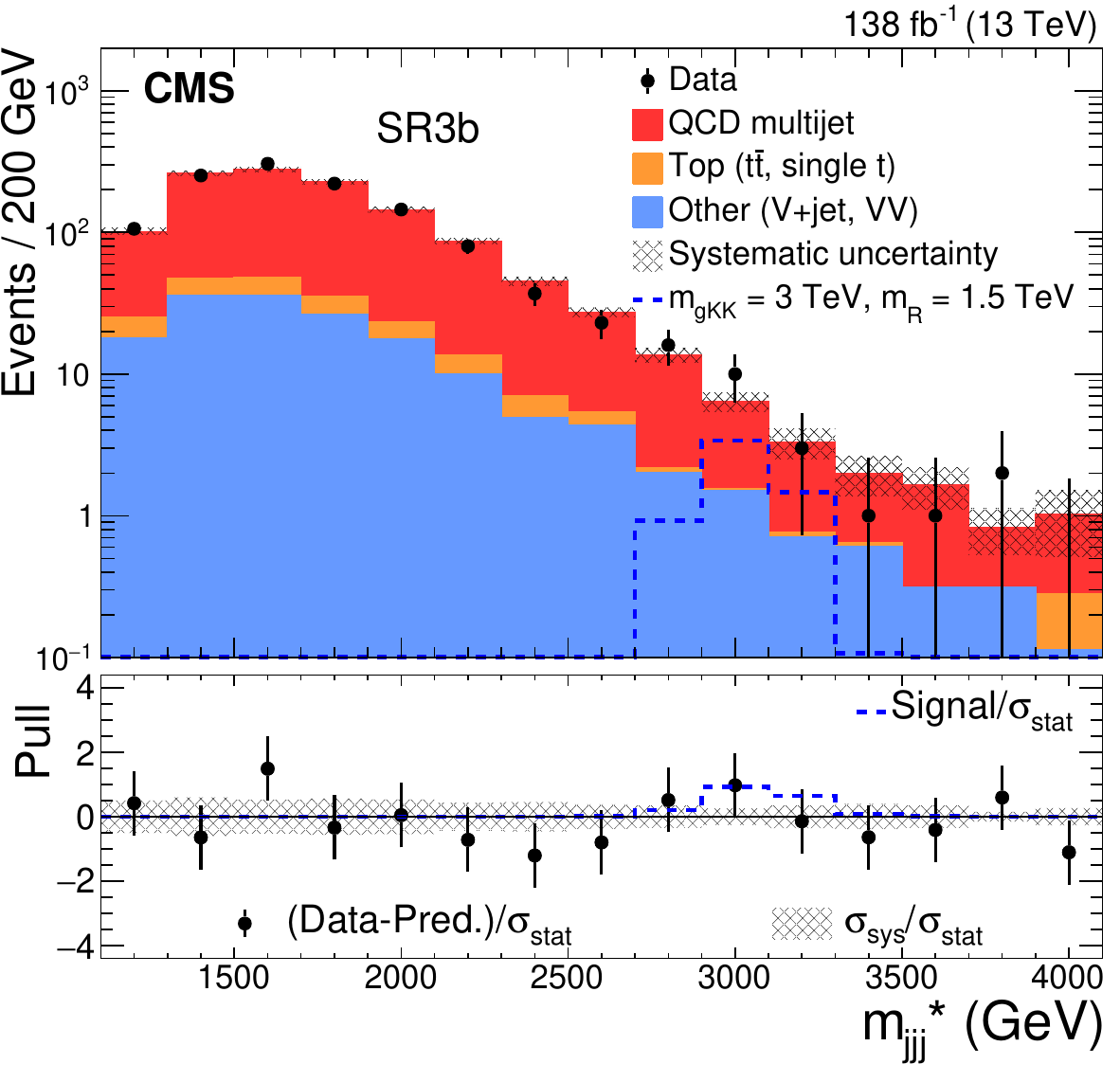}
\includegraphics[width=0.45\linewidth]{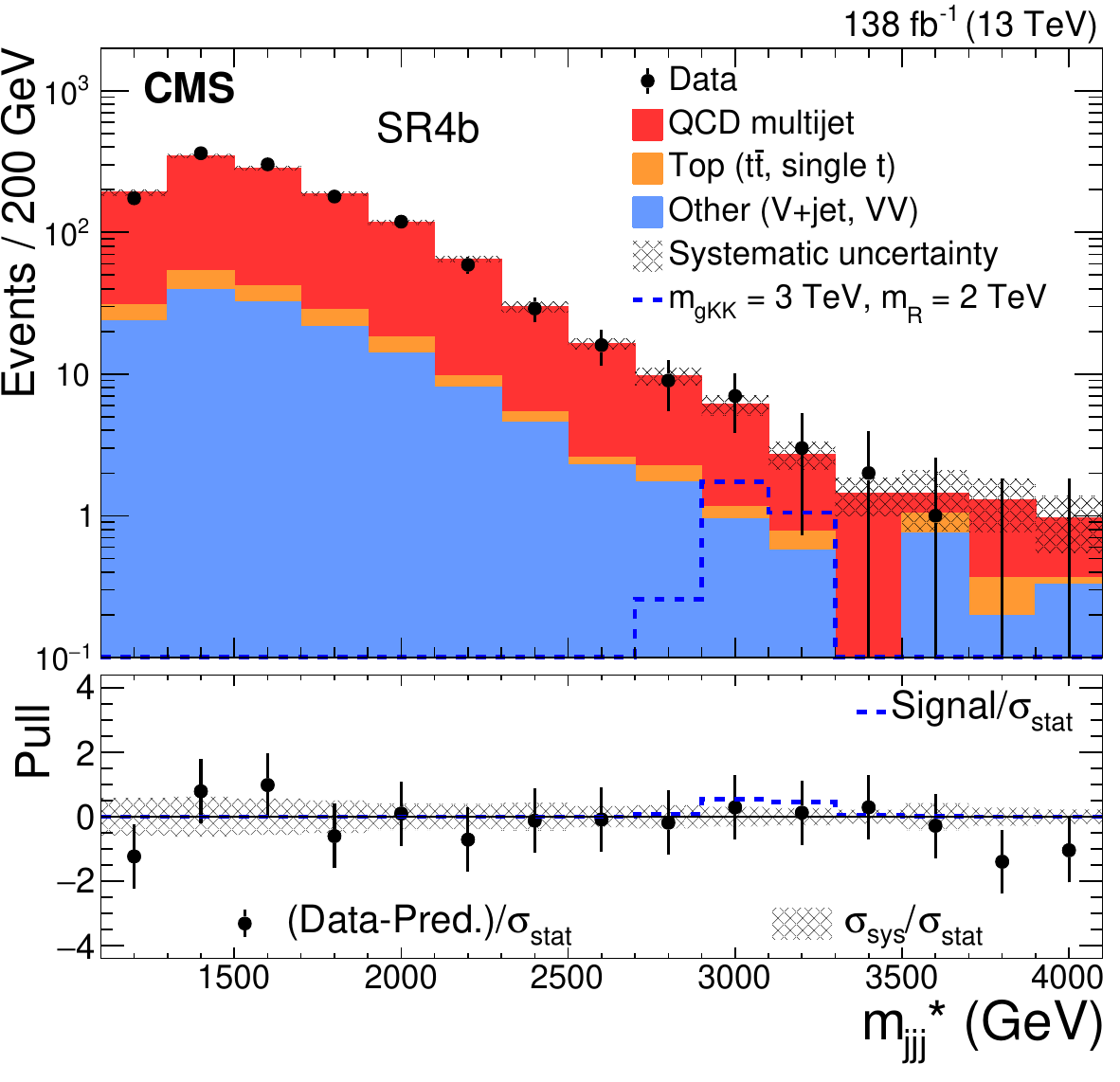}
\includegraphics[width=0.45\linewidth]{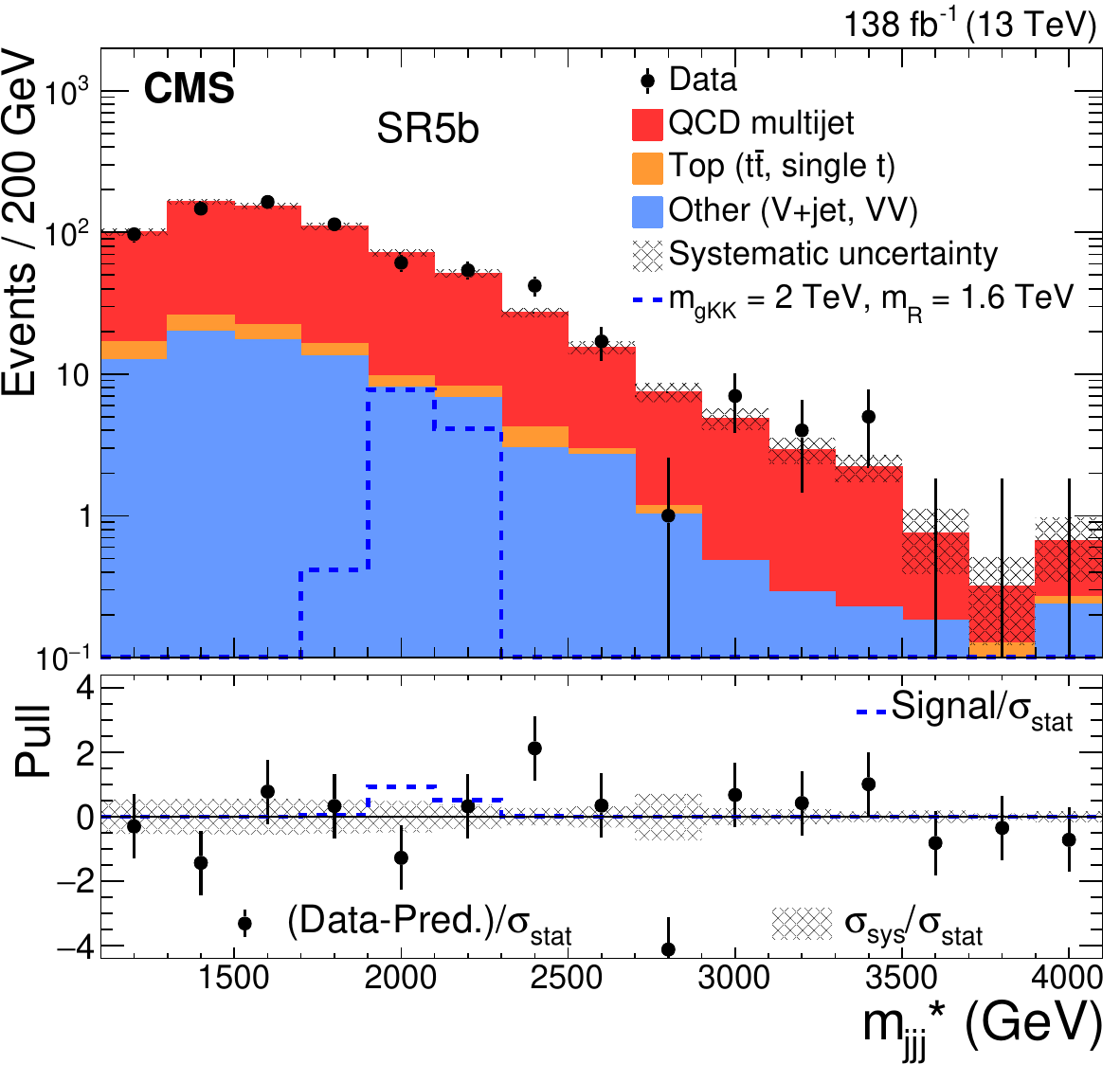}
\caption{The \mjjjs postfit spectra in the five SRxb.
Upper to lower, and left to right: SR1b, SR2b, SR3b, SR4b, and SR5b.
The lower panels show the pull distributions, as defined in the text, together with expected signal event yields and the systematic uncertainty normalized to $\sigma_{\text{stat}}$.}
\label{fig:Postfit_b}
\end{figure}

\subsubsection{Systematic uncertainties in the signal and background rates}

The uncertainties in the \PNetW calibration are presented in Table~\ref{tab:SFs}. 
A potential mismodeling of the \PW boson jet substructure dynamics, including the effects related to parton shower modeling, is corrected from the data via the SF's calibration.
The corresponding SF uncertainty, 6\% per jet matched to a \PW boson, accounts for residual jet shower mismodeling effects and the limited event sample used in the SF evaluation.
In case of events with two jets matching with \PW bosons, the uncertainty is taken as fully correlated, \eg{}, for a signal event in the SRa high-\pt bin, this results in a 12\% uncertainty.
These are considered for both signal and background processes that contain \PW-boson-matched jets.
The effect is accounted for by a nuisance parameter with a log-normal prior.

\section{Statistical analysis and results} \label{sec:Results}

The search for \gKK resonances is performed using a binned maximum likelihood fit to the \mjjjs distribution simultaneously in all ten SRs, as implemented with the CMS \textsc{Combine} tool~\cite{CMS:2024onh}.
The results are shown in Figs.~\ref{fig:Postfit_a} and \ref{fig:Postfit_b}.
The lower panels show the pull distributions defined as the difference between the data and predicted event yields divided by statistical uncertainty in the data, $\sigma_{\text{stat}}$.
Additionally, selected signal distributions and the total systematic uncertainty divided by $\sigma_{\text{stat}}$ are displayed.
The last bins include events beyond the shown range.
The data are in agreement with the estimated SM background within uncertainties.
The largest deviations between the data and the prediction are two single-bin downward fluctuations. These are located in SR3a and SR5b, $\mjjjs$ bin centers 3 and 2.8\TeV, with 0 and 1 data events observed for an expected 3.8 and 7.4 events, respectively.
They correspond to local \textit{p}-values of 2.1 and 0.5\%, assuming event counts with Poisson distributions.

For the background, the leading source of systematic uncertainty lies in the estimation of the QCD multijet background rates.
For the signal, the dominant uncertainty arises from the \PNetW selection efficiency calibration.
The overall result is statistically limited. 
This is presented at the lower panel of Figs.~\ref{fig:Postfit_a} and \ref{fig:Postfit_b} where the ratio of systematic to statistical uncertainties $\sigma_{\text{sys}}/\sigma_{\text{stat}}$ is less than 1.

\begin{figure}[!htbp]\centering
\includegraphics[width=0.65\linewidth]{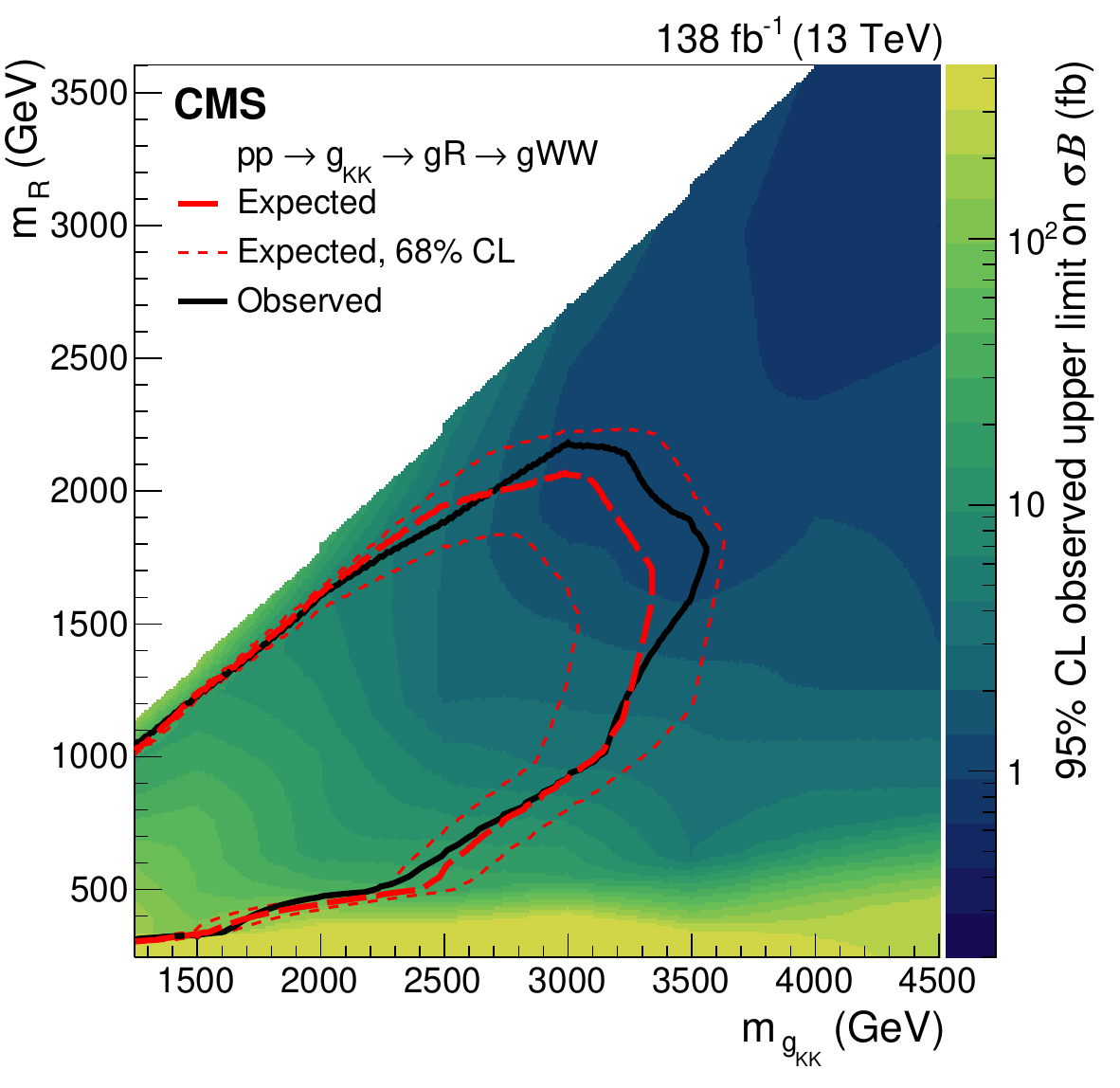}
\caption{The 95\% \CL upper limits on the production cross section times branching fraction $\sigma\mathcal{B}$, for the process $\Pp\Pp\to\gKK\to\Pg\PR\to\Pg\PW\PW$. The black solid (red dashed) line contour indicates the observed (expected) exclusion region of the \gKK and \PR masses. The red dotted lines show the 68\% \CL band of the expected limit contour.}
\label{fig:Limits}
\end{figure}

Upper limits at 95\% confidence level (\CL) are set on the production cross section times branching fraction for a potential resonance signal as functions of the \MgKK and \MR resonance masses, as shown in Fig.~\ref{fig:Limits}.
The limits are set following the modified frequentist approach 
and the definition of the profile likelihood test statistic as described in Ref.~\cite{CMS:2024onh} using an asymptotic approximation~\cite{Cowan:2010js}.

We exclude, at 95\% \CL, \gKK resonances decaying via a scalar radion \PR to a gluon and two \PW bosons, with \MgKK in the range of 1.25--3.55\TeV and \MR in the range of 0.3--2.18\TeV, using the model provided in Refs.~\cite{Agashe:2016rle,Agashe:2016kfr,Agashe:2017wss,Agashe:2018leo}.
The observed limits set in the \MgKK--\MR plane are stronger than the expected ones because of data under-fluctuations at $\mjjjs=3.5\TeV$ in SR3a and SR4a.

\section{Summary} \label{sec:sum}

A search for heavy resonances decaying, via an intermediate resonance, to a gluon and a pair of \PW bosons in the all-hadronic final states has been presented. 
A benchmark model, in which a Kaluza--Klein gluon decays into a radion and a gluon, $\gKK\to\Pg\PR\to\Pg\PW\PW$, has been considered.
The search is performed using proton-proton collision data at a center-of-mass energy of 13\TeV, collected by the CMS experiment, corresponding to a total integrated luminosity of 138\fbinv.
The final states include three large-radius jets, at least two of which are required to be massive, containing the decay products of the hadronically decaying \PW bosons.
The topology corresponds to events where each \PW boson from the \PR decay is reconstructed as a single merged jet.
In this analysis, a previously unexplored signature is probed using jet substructure techniques relying on deep learning.
No resonant signal above the standard model background prediction is found in the data analyzed.
Exclusion limits are set at 95\% confidence level on the masses of the \gKK and \PR bosons, up to 3.55 and 2.18\TeV, respectively, and on the product of the production cross section and the branching fraction to a gluon and two \PW bosons in an extended warped extra-dimensional model.
This result is the first analyzing the resonant $\PW\PW$+jet signature and complements constraints set on similar scenarios~\cite{CMS:2022lqh,CMS:2021mjl,CMS:2022tqn,CMS:2023tep}.

\begin{acknowledgments}
We congratulate our colleagues in the CERN accelerator departments for the excellent performance of the LHC and thank the technical and administrative staffs at CERN and at other CMS institutes for their contributions to the success of the CMS effort. In addition, we gratefully acknowledge the computing centers and personnel of the Worldwide LHC Computing Grid and other centers for delivering so effectively the computing infrastructure essential to our analyses. Finally, we acknowledge the enduring support for the construction and operation of the LHC, the CMS detector, and the supporting computing infrastructure provided by the following funding agencies: SC (Armenia), BMBWF and FWF (Austria); FNRS and FWO (Belgium); CNPq, CAPES, FAPERJ, FAPERGS, and FAPESP (Brazil); MES and BNSF (Bulgaria); CERN; CAS, MoST, and NSFC (China); MINCIENCIAS (Colombia); MSES and CSF (Croatia); RIF (Cyprus); SENESCYT (Ecuador); ERC PRG, RVTT3 and MoER TK202 (Estonia); Academy of Finland, MEC, and HIP (Finland); CEA and CNRS/IN2P3 (France); SRNSF (Georgia); BMBF, DFG, and HGF (Germany); GSRI (Greece); NKFIH (Hungary); DAE and DST (India); IPM (Iran); SFI (Ireland); INFN (Italy); MSIP and NRF (Republic of Korea); MES (Latvia); LMTLT (Lithuania); MOE and UM (Malaysia); BUAP, CINVESTAV, CONACYT, LNS, SEP, and UASLP-FAI (Mexico); MOS (Montenegro); MBIE (New Zealand); PAEC (Pakistan); MES and NSC (Poland); FCT (Portugal);  MESTD (Serbia); MCIN/AEI and PCTI (Spain); MOSTR (Sri Lanka); Swiss Funding Agencies (Switzerland); MST (Taipei); MHESI and NSTDA (Thailand); TUBITAK and TENMAK (Turkey); NASU (Ukraine); STFC (United Kingdom); DOE and NSF (USA).

\hyphenation{Rachada-pisek} Individuals have received support from the Marie-Curie program and the European Research Council and Horizon 2020 Grant, contract Nos.\ 675440, 724704, 752730, 758316, 765710, 824093, 101115353, 101002207, and COST Action CA16108 (European Union); the Leventis Foundation; the Alfred P.\ Sloan Foundation; the Alexander von Humboldt Foundation; the Science Committee, project no. 22rl-037 (Armenia); the Belgian Federal Science Policy Office; the Fonds pour la Formation \`a la Recherche dans l'Industrie et dans l'Agriculture (FRIA-Belgium); the F.R.S.-FNRS and FWO (Belgium) under the ``Excellence of Science -- EOS" -- be.h project n.\ 30820817; the Beijing Municipal Science \& Technology Commission, No. Z191100007219010 and Fundamental Research Funds for the Central Universities (China); the Ministry of Education, Youth and Sports (MEYS) of the Czech Republic; the Shota Rustaveli National Science Foundation, grant FR-22-985 (Georgia); the Deutsche Forschungsgemeinschaft (DFG), among others, under Germany's Excellence Strategy -- EXC 2121 ``Quantum Universe" -- 390833306, and under project number 400140256 - GRK2497; the Hellenic Foundation for Research and Innovation (HFRI), Project Number 2288 (Greece); the Hungarian Academy of Sciences, the New National Excellence Program - \'UNKP, the NKFIH research grants K 131991, K 133046, K 138136, K 143460, K 143477, K 146913, K 146914, K 147048, 2020-2.2.1-ED-2021-00181, and TKP2021-NKTA-64 (Hungary); the Council of Science and Industrial Research, India; ICSC -- National Research Center for High Performance Computing, Big Data and Quantum Computing and FAIR -- Future Artificial Intelligence Research, funded by the NextGenerationEU program (Italy); the Latvian Council of Science; the Ministry of Education and Science, project no. 2022/WK/14, and the National Science Center, contracts Opus 2021/41/B/ST2/01369 and 2021/43/B/ST2/01552 (Poland); the Funda\c{c}\~ao para a Ci\^encia e a Tecnologia, grant CEECIND/01334/2018 (Portugal); the National Priorities Research Program by Qatar National Research Fund; MCIN/AEI/10.13039/501100011033, ERDF ``a way of making Europe", and the Programa Estatal de Fomento de la Investigaci{\'o}n Cient{\'i}fica y T{\'e}cnica de Excelencia Mar\'{\i}a de Maeztu, grant MDM-2017-0765 and Programa Severo Ochoa del Principado de Asturias (Spain); the Chulalongkorn Academic into Its 2nd Century Project Advancement Project, and the National Science, Research and Innovation Fund via the Program Management Unit for Human Resources \& Institutional Development, Research and Innovation, grant B39G670016 (Thailand); the Kavli Foundation; the Nvidia Corporation; the SuperMicro Corporation; the Welch Foundation, contract C-1845; and the Weston Havens Foundation (USA).
\end{acknowledgments}
\bibliography{auto_generated} 

\cleardoublepage \appendix\section{The CMS Collaboration \label{app:collab}}\begin{sloppypar}\hyphenpenalty=5000\widowpenalty=500\clubpenalty=5000
\cmsinstitute{Yerevan Physics Institute, Yerevan, Armenia}
{\tolerance=6000
A.~Hayrapetyan, A.~Tumasyan\cmsAuthorMark{1}\cmsorcid{0009-0000-0684-6742}
\par}
\cmsinstitute{Institut f\"{u}r Hochenergiephysik, Vienna, Austria}
{\tolerance=6000
W.~Adam\cmsorcid{0000-0001-9099-4341}, J.W.~Andrejkovic, L.~Benato\cmsorcid{0000-0001-5135-7489}, T.~Bergauer\cmsorcid{0000-0002-5786-0293}, S.~Chatterjee\cmsorcid{0000-0003-2660-0349}, K.~Damanakis\cmsorcid{0000-0001-5389-2872}, M.~Dragicevic\cmsorcid{0000-0003-1967-6783}, P.S.~Hussain\cmsorcid{0000-0002-4825-5278}, M.~Jeitler\cmsAuthorMark{2}\cmsorcid{0000-0002-5141-9560}, N.~Krammer\cmsorcid{0000-0002-0548-0985}, A.~Li\cmsorcid{0000-0002-4547-116X}, D.~Liko\cmsorcid{0000-0002-3380-473X}, I.~Mikulec\cmsorcid{0000-0003-0385-2746}, J.~Schieck\cmsAuthorMark{2}\cmsorcid{0000-0002-1058-8093}, R.~Sch\"{o}fbeck\cmsAuthorMark{2}\cmsorcid{0000-0002-2332-8784}, D.~Schwarz\cmsorcid{0000-0002-3821-7331}, M.~Sonawane\cmsorcid{0000-0003-0510-7010}, W.~Waltenberger\cmsorcid{0000-0002-6215-7228}, C.-E.~Wulz\cmsAuthorMark{2}\cmsorcid{0000-0001-9226-5812}
% ...
\par}
\vskip\cmsinstskip
\dag:~Deceased\\
$^{1}$Also at Yerevan State University, Yerevan, Armenia\\
$^{2}$Also at TU Wien, Vienna, Austria\\
% ...
\end{sloppypar}
\end{document}